\documentclass[aps,prb,twocolumn,groupedaddress,showpacs,showkeys,floatfix,]{revtex4}
\usepackage{amsmath}
\usepackage{amsfonts}
\usepackage{amssymb}
\usepackage{graphicx}
\include{psfig}
\preprint{}

\begin{document}
\title{\bf
Critical Roles of Metal-Molecule Contacts in Electron Transport Through Molecular-wire Junctions}
\author{
A. Grigoriev, J. Sk\"oldberg, G. Wendin}
\affiliation{Department of Microtechnology and Nanoscience - MC2, \\
Chalmers University of Technology, SE-41296 Gothenburg, Sweden}
\author{
\v Z. Crljen}
\affiliation{R. Boskovic Institute, P.O. Box 180, HR-10002 Zagreb, Croatia}

\date{\today}

\begin{abstract}
We study the variation of electron transmission through Au-S-benzene-S-Au junctions and related systems as a function of the structure of the Au:S contacts.
For junctions with semi-infinite flat Au(111) electrodes, the highly coordinated in-hollow and bridge positions are connected with broad transmission peaks around the Fermi level, due to a broad range of transmission angles from transverse motion, resulting in high conductivity and weak dependence on geometrical variations. 
In contrast, for (unstable) S-adsorption on top of an Au atom, or in the hollow of a 3-Au-atom island, the transmission peaks narrow up due to suppression of large transmission angles. Such more one-dimensional situations may describe more common types of contacts and junctions, resulting in large variations in conductivity and sensitivity to bonding sites, tilting and gating.
In particular, if S is adsorbed in an Au vacancy, sharp spectral  features appear near the Fermi level due to essential changes of  the level structure and hybridization in the contacts, admitting order-of-magnitude variations of the conductivity. 
Possibly such a system, can it be fabricated, will show extremely strong non-linear effects and might work as uni- or bi-directional voltage-controlled 2-terminal switches and non-linear mixing elements. 
Finally, density-functional-theory (DFT) based transport calculations seem relevant, being capable of describing a wide range of transmission peak structures and conductivities. Prediction and interpretation of experimental results probably require more precise modeling of realistic experimental situations.

\end{abstract}

\pacs{71.15.Mb; 68.35.Bs; 68.47.De}
\keywords{Molecular electronics; Nonequilibrium electron transport: Density functional calculations; Organic molecules; Nanostructures; Gold}

\maketitle

\section{\label{Introduction}Introduction}

Owing to their stability, thiol-terminated organic molecules bridging gold electrodes have been attractive systems for investigating transport properties of molecular junctions \cite{Hipps2001,NitzanRatner2003}. In particular, the dithiol-benzene (DTB) molecule  coupled to Au(111) gold surfaces \cite{Reed1997} has been a kind of model system for  theoretical studies. \cite{Larsson,Stokbro2003,Yaliraki1999,DiVentra2000,Derosa2001,Xue2001,Grigoriev2003, MRS2003, Bratkovsky2003,Bauschlicher2003,Evers2003,Basch2005,WangFuLuo2005}

After the first semi-empirical investigations, \cite{Yaliraki1999,Emberly1998,Emberly2001} calculations within the framework of density functional theory (DFT) have been performed on DTB \cite{DiVentra2000,Derosa2001,Xue2001,Grigoriev2003,MRS2003,Bratkovsky2003,Bauschlicher2003,Evers2003,Basch2005,WangFuLuo2005,Ke2005a,Ke2005b,ThygeJacob2005} and similar systems \cite{Shi2003,Yang2003}. The conductance was found to be more than an order of magnitude larger than experiment, despite assumptions of unequal bonding of the molecule to the electrode surfaces.\cite{Stokbro2003}
The difference was related to varying strengths of the sulfur-gold bonds and to the adsorption sites

The transmission properties of the systems are determined by the electronic structure of the combined molecule and electrodes systems, the metal-molecule-metal (M-mol-M) junction. The direct tunneling between electrodes, in the absence of molecules, is small for the typical size of the molecular junctions. 
The distance between electrodes has an important role in the conduction of the junction, and its variation is strongly correlated with the changes in atomic and electronic structure of the extended molecule.. 

With the molecule bridging the interelectrode gap, the transport through the open system is determined by scattering states which are Bloch-like in the electrodes and molecular-like in between the electrodes.
However, these extended states can have very different weights, some being localized to the contacts, some on the bridging molecule, and some delocalized over the entire M-mol-M junction. We have recently \cite{Crljen2005}  investigated the transport properties of the thiol-terminated OPVn, n=3,4,5 series of molecular wires, focussing on the length dependence for fixed structure of the M-mol contacts, i.e. Au-S adsorbate site and level structure. 

In this paper we focus on metal-molecule contacts, and investigate the Au(111) adsorption-site dependence of the transmission for DTB and several other short S-terminated molecules. The purpose is to see if there are situations with low transmission that can explain the discrepancy between theory and experiment. Some preliminary results have ben presented in Refs.\cite{Grigoriev2003,MRS2003}.

In Secs. II and III of the paper we first describe the model used and the generalites of the computational procedure. In Sec. IV we describe and compare the basic features of the zero-bias transmission spectra.  In Sect. V we calculate the current through the system as a function of applied bias voltages and analyze them in terms of the evolution of transition spectra under bias. Section VI discusses some aspects of resonances with Fano line shapes, and Sec. VII provides some concluding remarks.

\section{\label{Molecules} Molecules and junctions}

The configuration of the electrode-molecule-electrode system is crucially important for the calculating realistic transmission spectra. In this work we study the dependence of the transmission spectra primarily at zero bias voltage (i.e. zero bias conductance) on the adsorption site on 111 gold surfaces. In our calculations, we primarily study the molecules shown in Fig.1, with thiolate bonds S:Au(111) for sulfur adsorbed on gold electrodes. 
We find that for 1,4-Benzenedithiol (DTB) on the flat 111 gold surface at $\theta=1/9$ coverage the dependence on the adsorption site and on the rotation of the benzene ring is negligible. However, in the presence of vacancies (missing surface gold atoms) the transmission spectrum and the conductance change significantly. We compare the latter result with the Butadiene-1,4-dithiol (DTBd), 1,4-Benzenedimethanethiol (DMtB) and 1,3- 1,4-di thiol bicyclo[2.2.2]octa-2,5,7-triene (DTBO) transmission data.

A typical molecular electronic system consists of a molecule coupled to two electrodes with different electro-chemical potentials. Depending on the experimental method of building the device the contact geometry may vary with respect to adsorption site as well as to the adsorption strength. 

\begin{figure}
\centerline{
\includegraphics[width=7cm]{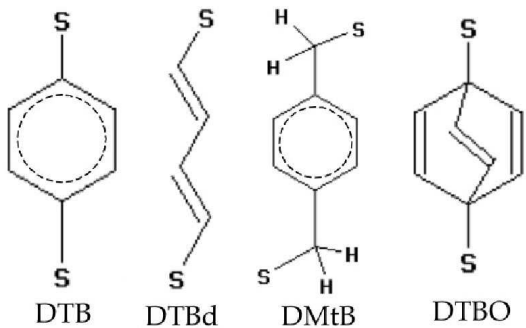}
}
\caption{The four molecules considered in this work, connected to two Au(111) surfaces via thiolate or thiol bonds. From left to right: DTB, DTBd, DMtB and DTBO. The S-S distances are (H-terminated/sandwiched between Au(111) electrodes) $6.40/6.37 A$, $6.85/6.89 A$, $8.14/7.997 A$ and $6.25/6.23 A$ resp. (Without termination, S-B-S is $6.15A$ and S-CH$_2$-B-CH$_2$-S is  $6.19A$).
}
\label{4molecules}
\end{figure}
\noindent

\begin{figure}
\centerline{
\includegraphics[width=8cm]{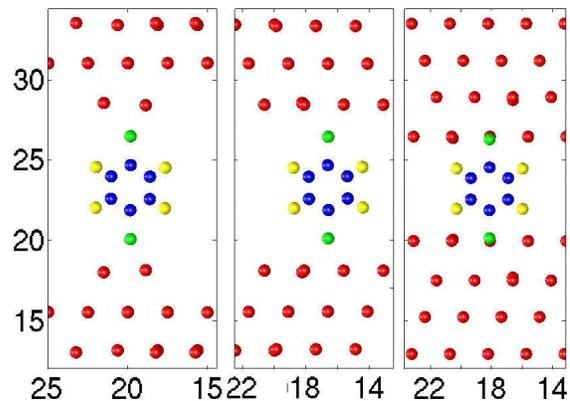}
}
\caption{Au-S-benzene-S-Au junctions with three types of contacts: Left: S in the hollow of a 3-Au atom island ("point contact"). Middle: S in hcp (3-fold hollow) site of relaxed Au(111). Right: S in an Au vacancy. The distances are in Angstroms.  The distances between the first surface layers of electrodes are given in Table 2 (Fig.7).
}
\label{DTB_sites}
\end{figure}

We consider three distinct types of Au electrode surfaces and Au-S junction contacts, as shown in Fig. \ref{DTB_sites},
namely (1) S in the hollow of a 3-Au-atom island ("point contact") (left panel); (2) S adsorbed on a relaxed Au(111) surface (middle panel); (3) S buried in Au vacancy (right panel).

\begin{figure}
\centerline{
\includegraphics[width=7cm]{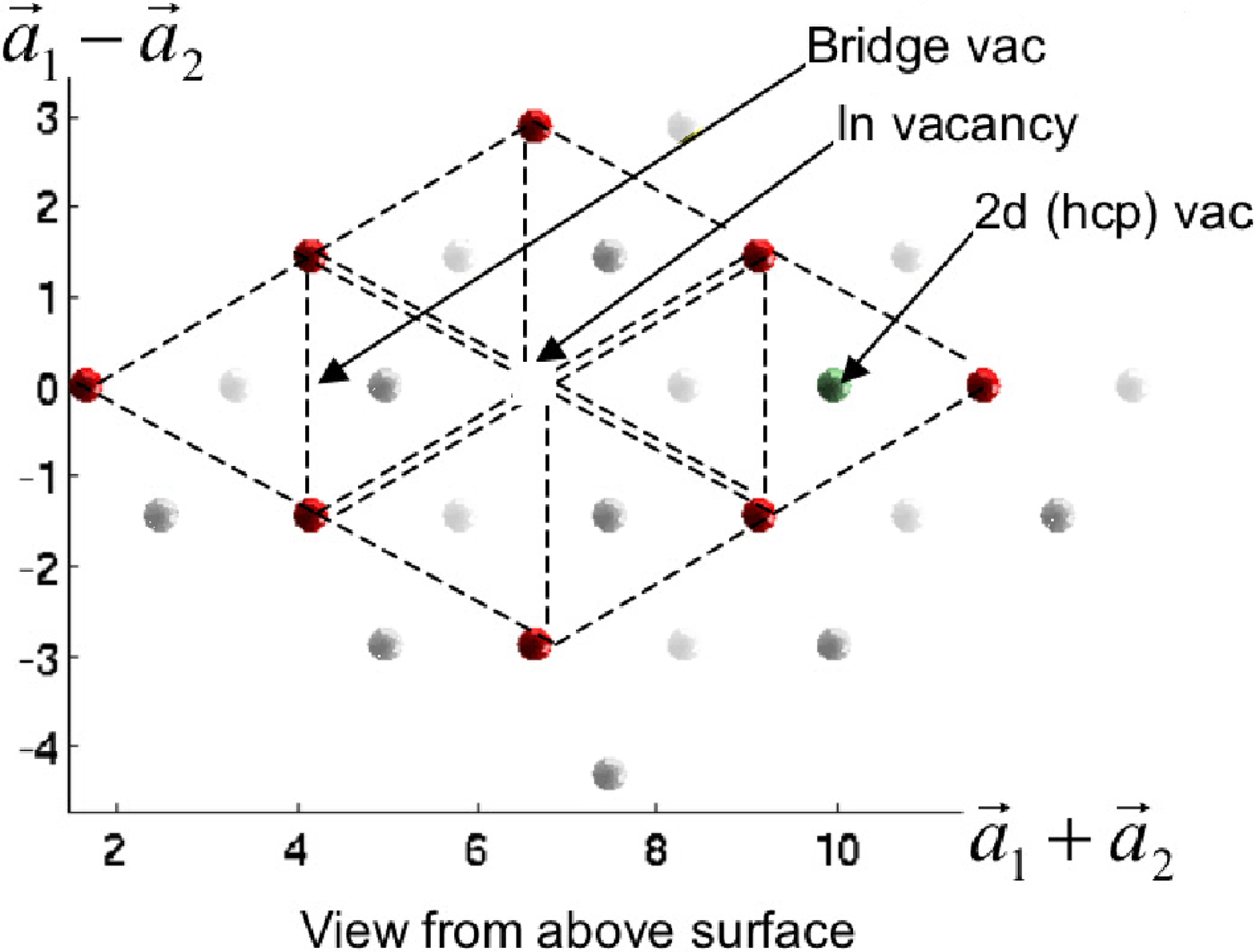}
}
\vspace{0.5cm}
\centerline{
\includegraphics[width=7cm]{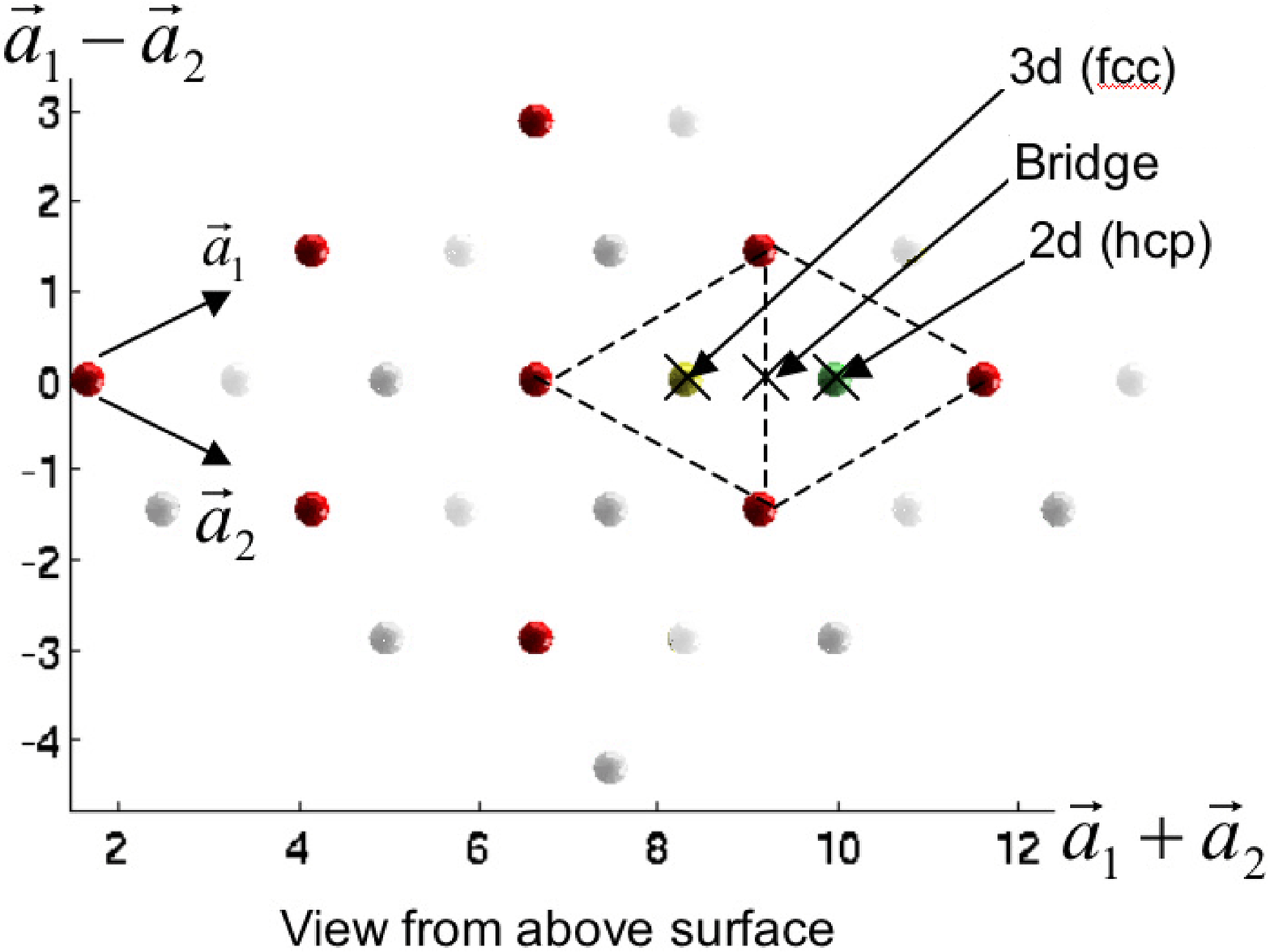}
}
\caption{Definition of adsorption sites.  (3d(fcc) denotes a site on top of a gold atom in the 3rd layer; 2d (hcp) denotes a site on top of a gold atom in the 2nd layer).}
\label{Sites}
\end{figure}

For relaxed Au(111) electrode surfaces we consider a number of S-adsorption sites shown in Fig.\ref{Sites}, top panel: hcp 3-fold hollow site, bridge site, and on-top position. In addition (Fig.\ref{Sites}, bottom panel), we also consider the hcp, fcc and bridge sites  near a vacancy. 
Note that several aspects unavoidably get mixed in such a comparative study. In addition to probing different types of Au-S contacts, the distance between the Au-planes and effective Au-S contacts vary, which may significantly influence the hybridization of the Au electrodes and the transport properties.

\section{Computational methods and results}
\subsection{Theoretical background}
In order to perform first principles quantum modeling of electronic structure under non-equilibrium conditions and to calculate the IVCs, we use non-equilibrium Greens function (NEGF) techniques based on density functional theory (DFT) for the electronic structure, as implemented in the TranSIESTA simulation package \cite{Brandbyge2002}.
Core electrons were modeled with Troullier-Martins \cite{TMpseudopot1993} soft norm-conserving pseudopotentials and the valence electrons were expanded in a basis set of local orbitals \cite{SIESTA}.
The system is divided into three regions: left and right bulk electrodes and central region. The central part contains the portion of physical electrodes where all the screening effects take place. The Fermi level of the overall system is determined by the charge neutrality of the central region. The charge distribution in the electrodes corresponds to the bulk phases of the same material.

The density matrix of the system under the external bias is calculated self-consistently within DFT using the standard local-density approximation (LDA) for the exchange correlation functional. The virtue of the method is that it gives the Hamiltonian in the same form as in the empirical tight-binding approach, making its techniques straightforwardly applicable.

An important aspect of the calculation is the robustness of the results to
computational details.
From the calculation by Stokbro {\em et al.} \cite{Stokbro2003} for the DTB on Au(111) in the hollow position we know that the use of the generalized gradient approximation (GGA) instead of the LDA gives almost negligible effects on the transmission amplitude. The insensitivity of transport results to (non-) relaxation of the first two gold layers was also demonstrated \cite{Stokbro2003}. As the binding geometry is mostly determined in the first two layers of gold atoms, the conclusion is valid for systems with other molecules as well.

\subsection{Electronic structure calculations}

\subsubsection {Adsorption on Au(111) surface}
We use $15$ k-points in $551$ Monkhorst Pack scheme for $2\times2$ supercell and $6$ k-points in $331$ Monkhorst Pack scheme for $3\times3$. Following the TranSIESTA setup \cite{SIESTA}, we use minimal basis set for gold species. The LDA approximation for exchange-correlation functional is used, since the LDA results for gold surfaces agree well with experiment \cite{Mankefors2003} and is used in the TranSIESTA setup.

\begin{figure}
\centerline{
\includegraphics[width=7cm]{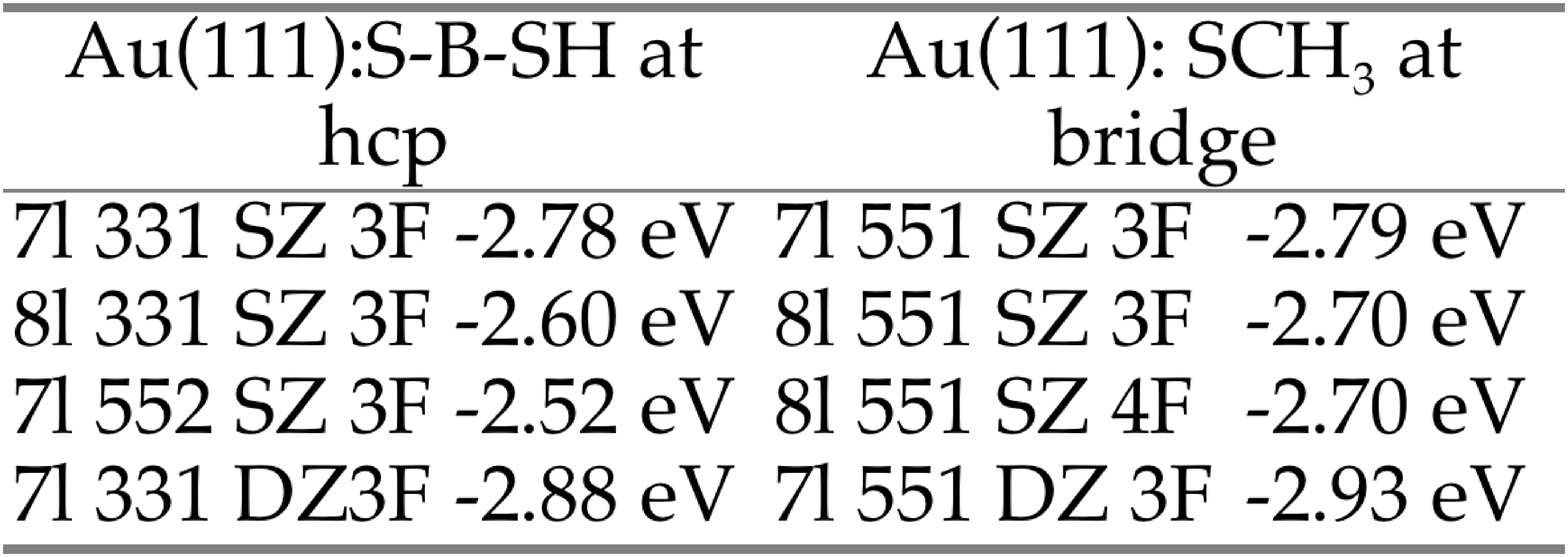}
}
\caption{TABLE 1 - Convergence tests for the number of layers of gold, k-points sampling (Monkhorst-Pack scheme), basis size for Au and number of fixed "bulk" layers. Coverage is $1/9$ for $Au(111):S-B-SH$ and $1/4$ for $Au(111): SCH_3$.}
\label{table1}
\end{figure}

The gold surface is modeled by a $7$-layer gold slab in $111$ stacking with $3$ layers of gold atoms fixed in bulk positions at one side and $4$ layers allowed to fully relax to model the free surface at the other side. (It is known \cite{Yourdshahyan2001} that to correctly reproduce the surface energy of the Au(111) surface with 6 k-points, minimum 5 layers of gold is required.)
Since in metals the disturbances of charge density caused by surfaces or defects in crystal structure are effectively screened and do not spread over more than about $3$ layers, the free surface behave as a surface of bulky metal and not as a second surface of a thin metal slab. The details follow our previous work \cite{Mankefors2003}; however, since the size of the studied adsorbate is much bigger, the thickness of the slab was increased to ensure, that the charge transfer between the adsorbed molecule and the surface is realistically modeled. Direct subtraction of the bulk charge density shows that there is no difference between bulk and the electrode-part of the central region of the slab. The convergence test (Fig.\ref{table1}) with respect to the number of layers was performed with the SCH$_3$ molecule in bridge position ($\theta =1/4$); no significant change was observed for an $8$-layer slab.

Our results for SCH$_3$ adsorption qualitatively reproduce results from Ref.\cite{Yourdshahyan2001}, however the energy differences for the $7$ layer slab setup are about $10$ times smaller, i.e. the adsorption energy for SCH$_3$ at the hcp site is $-2.842\; eV$ compared to $-2.787\; eV$ at the bridge site. The discrepancy with Refs.\cite{Hyashi2001,Gronbeck2000} can be easily explained by the difference in the computational setup. The $4$-layer gold slabs used in Refs.\cite{Hyashi2001,Gronbeck2000} correspond to a situation with adsorption on a thin metal film, rather than on an electrode surface. Additional discrepancies with Ref.\cite{Gronbeck2000} and with analogous results for DTB \cite{Nara2004} are attributed to the number of gold layers with relaxed atomic positions (one in Refs.\cite{Gronbeck2000,Nara2004}). The number of layers used in the calculation is especially important, since $S$-terminated aromatic molecules on thin gold films ($3$ layers, atoms constrained to bulk positions) prefer top and bridge sites \cite{Krstic2003}.

Our primary goal is to achieve sufficient accuracy for reliable calculation of transport properties, i.e. the main features in transmission spectra, IVCs and differential conductivities. We found, that a) although the energy is stabilized upon adding the $8$th layer of gold atoms, some of the gold atoms inside the surface layers move up to 6 percent of the interatomic distance and b) our setup does not fully reproduce the "nobility" of gold, namely the adsorption energy of the H atom from H2 molecule is negative. This could be attributed to the insufficient number of k-points and the small size of the basis used for gold. However, we have tested the influence of these errors on the transmission spectra. We find that the change in transmission due to the change of the basis size for the surface gold atoms, or the minor distortion of the structure, is of the order of $0.05$. 

$9$ gold atoms ($3\times3$ surface cell) were used in each layer for $\theta=1/9$ coverage, periodically repeated in space. The molecular adsorption was studied by allowing full relaxation of the atomic coordinates of the molecule and the 4 topmost surface layers. The adsorption energy was calculated according to 

\begin{equation}
E_a = E_{Au:M} - E_{Au} - E_{M}
\end{equation}
where  M stands for the molecule at study. If we assume, that prior to adsorption the sulfur was terminated with an H atom, then the reaction path would look like

\begin{equation}
HM + Au \rightarrow Au:M + H_2/2 \;,
\end{equation}
yielding the energy balance 

\begin{equation}
\Delta E = E_{Au:M} + E_{H_2}/2- E_{Au} - E_{HM} \;.
\end{equation}
We emphasize that the adsorption energy calculated along this path will be different by

\begin{equation}
\Delta E-E_a = E_M - E_{HM} + E_{H_2}/2 \;,
\end{equation}
which is constant for a given molecule in a given cell and for a given simulation setup. 
For the typical setup, the correction is $1.47\;eV$ for DTB, $1.35\;eV$ for DTBd, $1.92\;eV$ for DTBO and $1.93\;eV$ for DMtB.
The $3\times3$ surface cell setup for gold does not provide space enough for a realistic study of the molecular film structure. Molecules in the film are expected to adsorb at two different angles \cite{Dhirani1996,Himmel1998,Sabatani1993,Nara2004,Krstic2003}. However, for the DTB molecule it is possible to study rotation, in the sense that all the molecules in the film stay at the same angle and even when the benzene ring turns along the shortest cross-section of the supercell it does not "touch" the nearest neighbor. For this study we used the same setup as before, but the relaxation every time starting from the configuration relaxed in the previous run with the benzene ring rotated around the S-S axes by the angle $\alpha$, where $\alpha=0$ was chosen to be with the ring in the $(1\bar1 0)$ plane. 
The adsorption energy changes upon rotation are within +/- 0.05 eV.   \\

We should either say here that the changes in transmission are small or discuss this issue later. \\

\subsubsection {Metal-molecule-metal sandwich}
For the M-mol-M sandwich setup, the metal electrode separation is an external parameter that plays a crucial role. Depending on the experimental conditions, the distance between the metal electrodes can be determined by fixed (or variable)-geometry metal contacts, or by the molecular layer itself, with added metal contacts. For the theoretical calculations to model these different experimental situations, different initial conditions may be appropriate.

For a molecular monolayer formed on the surface, it is natural to start from a molecular adsorption calculation and proceed by adding the second electrode symmetrically, i.e. starting the sandwich relaxation from the relaxed monolayer interfaces (our choice). For molecules deposited in the gap between electrodes, the natural choice could be the free molecule between two free surfaces, giving (for the same gold - sulfur distance) different distances between gold bulk regions, and, after the relaxation of the structure, different interface geometries. 

Making mechanical contact (like in break junctions, or in the method of gold cluster deposition) to a molecular film when the film is formed on one electrode and another electrode is initially clean, leaves less freedom in controlling the electrode separation: the formation of the chemical bond to the second electrode takes place at some particular separation and can be sensed through the I-V regime of the device, since before the chemical bond is established, the expected I-V is asymmetric. However, when the molecule is chemically bonded to both contacts, the force on the molecule is controlled by the break junction or by AFM contact to the gold cluster \cite{Hipps2001,NitzanRatner2003}.

The two possible distinct positions for the device operation are the one just after the contact is established and the one after which the contact is known to break. Since the bond to the surface is energetically favorable, such a point for the nearly upright-standing molecules in the film will be on the maximum reach from the first electrode, i.e. when the sulfur to sulfur vector is perpendicular to the gold surface. We have chosen to model this experimental situation, when the chemical bond is symmetrically established to both electrodes, the molecule is upright and the strain in the transport direction is minimized. The situation when the film is stretched or compressed, and when the film is tilted by applying the lateral force to the contact will be discussed elsewhere.

The chosen setup makes it possible to solve the geometrical problem and use symmetric electrodes with fcc stacking ...ABCABC-CBACBA... in the transport direction. Alternative stacking and a configuration with stacking fault were tested with no significant effect on the transmission.

\begin{figure}
\centerline{
\includegraphics[width=8.5cm]{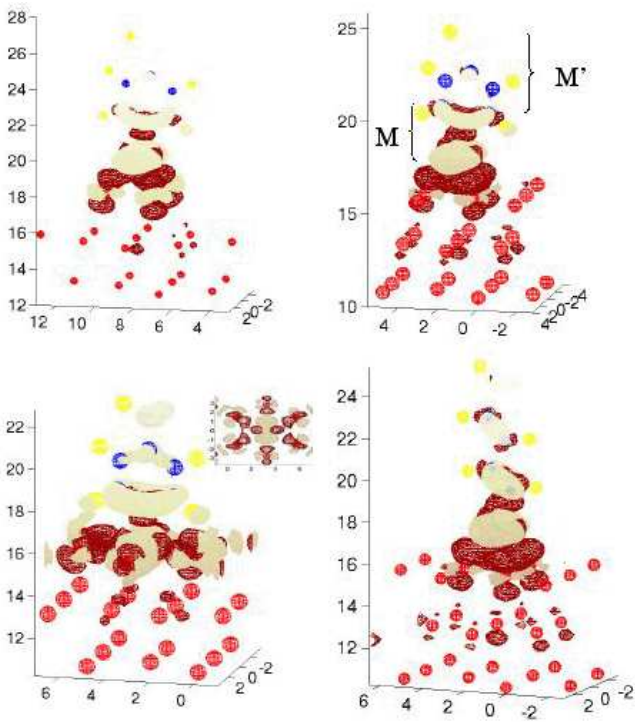}
}
\caption{Charge redistribution for the bond formation. Top row: Left: DTB on three adatoms at hcp; right: on flat surface hcp. Bottom row: Left: DTB in vacancy; right: DTBd at hcp. Isosurfaces enclose regions with maximum charge redistribution and contain 40 percent of the displaced (solid) or removed (facet) charge. The distances are in Angstroms. Inset shows the view of the surface layer with S atom in vacancy. The notation M and MÕ is discussed in the text.
}
\label{DTB_contacts}
\end{figure}

We adopted the following algorithm of creating the sandwich out of the symmetric molecule adsorbed on the surface: for the Au:S-MM'-SH system, where M and M' stands for the two halves of the benzene ring in the case of DTB (see Fig.\ref{DTB_contacts}) , we found, that charge redistribution and change of geometry upon adsorption takes place predominantly along the S-M part. Hence, the Au:S-MMir-S:Aui geometry was adopted for the sandwich setup, where i denotes inversion of z coordinates and r - (-rotation around z  axis. We tested this geometry in a periodic supercell with 9 layer gold slab, and found the relaxation to be negligible and not influencing the transmission results. This operation greatly speeds up the calculation, since the relaxation of the sandwiched molecule requires large resources. Although, this algorithm works fine for the planar molecules, it is inapplicable to the 3D shaped molecules, and we used relaxation in the supercell geometry in those cases. 
\begin{figure}
\centerline{
\includegraphics[width=8cm]{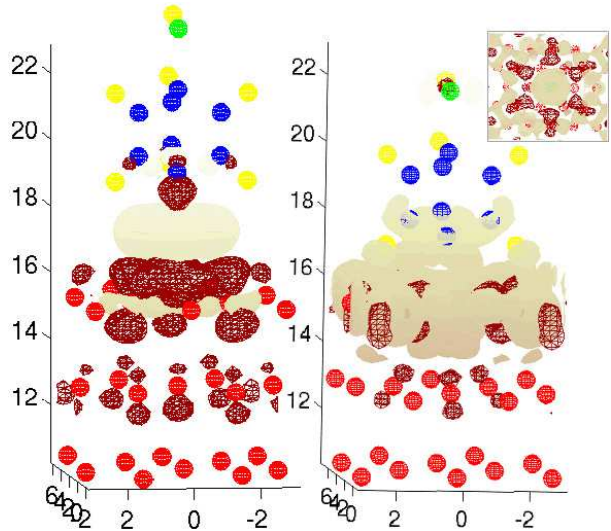}
}
\caption{Charge redistribution for the bond formation. DTBO at hcp and in vacancy. Isosurfaces enclose regions with maximum charge redistribution and contain 40 percent of the displaced (solid) or removed (facet) charge. The distances are in Angstroms. Inset shows the view of the surface layer with S atom in vacancy.}
\label{DTBO_contacts}
\end{figure}

\begin{figure}
\centerline{
\includegraphics[width=8cm]{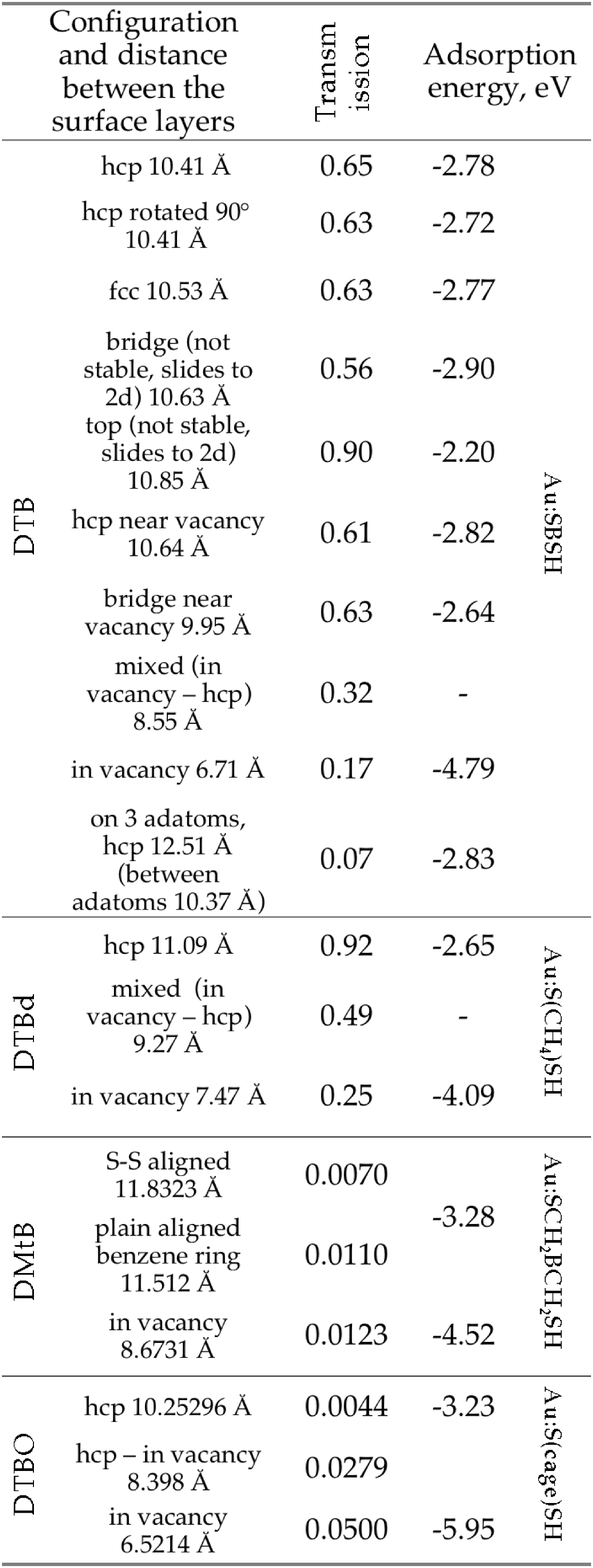}
}
\caption{
Table 2: Transmission at the Fermi level for the four molecules considered in this work for different adsorption sites.
}
\label{Table2}
\end{figure}

Along with the expected dependence on the molecule, Table 2 (Fig.\ref{Table2}) demonstrates the pronounced dependence of the transmission on the sulfur back bonding. The transmission drops up to two orders of magnitude, when the carbon atom connected to the sulfur has its maximum coordination number. The influence of the specific back-bonding on the structure of the adsorbate has already been studied \cite{Majumder2002}, however we are not aware of the systematic data on the back-bonding effect on the molecular transmission. In Fig.\ref{DTB_contacts} we present the adsorbed molecules at 1/9th  coverage, and charge difference due to the charge redistribution upon the bond formation between the molecule and the gold. Charge redistribution due to the atomic relaxation is not shown. Both in vacancy and adsorption on the flat surface (hcp site) is considered.
As already mentioned, the charge is mainly redistributed around the S atom and it's nearest neighbors. The only exception is DTBd with it's linear structure. Another important feature common for all the presented molecules is that charge is mainly pulled from the gold towards the molecule, but only surface gold atoms are affected, even if the molecule sits in vacancy. However, the bond is not completely ionic, since the charge is not only moved towards sulfur, but also is redistributed around the sulfur and between sulfur and gold, indicating the covalent character of the bond. This is especially true for the in-vacancy adsorption of the DTBO, Fig.\ref{DTBO_contacts}. The charge build up around the surface gold atoms indicates the interaction with the d-levels of gold.
The two types of back-bonding influence can be distinguished by the charge accumulation around the sulfur. The filled figure-eight-shaped lobe (Fig.\ref{DTB_contacts}) indicates that the $\pi$-type orbital is filled when the carbon's coordination is 3 and the torus lobe (Fig.\ref{DTBO_contacts}) is filled in the other case. The same type can be found for the SCH$_3$ adsorption. The latter type has clearly poor overlap with the $\pi$-type orbital structure on the benzene ring and somewhat better overlap with the $\pi$-type orbitals around the carbon-carbon double bonds in the DTBO, because the latter are respectively rotated by $60\deg$. This in part explains the differences in the transmission magnitude. 
For the sandwiched DTB molecule the Mulliken charges on the S atoms are 6.391 and 6.388 at the hcp site and 6.343 and 6.339 in the vacancy.

There is an obvious correlation between the chemistry of the interface and the charge redistribution. For the DTB and DTBd molecules the thiol group is connected to the $\pi$-bonded complex, while it is not the case for the DTBO and DMtB, and the charge redistribution upon adsorption reflects this bonding character. Charge redistribution upon adsorption was found to be about 0.1 percent of the total pseudo charge of the system.

Table 2 (Fig.\ref{Table2}) collects the results for the transmission at the Fermi level for the various M-mol-M systems and adsorption sites considered. In the cased of DTB, one notes that the variations on the flat relaxed Au(111) are small or insignificant, even in the case of a nearby vacancy. Considerable variations are obtained only when the sulfur head is buried in a  gold vacancy, or adsorbed on a small island. However, as we will see, to get good understanding it is important to study the transmission spectrum over a wider range of energies, because the spectral changes can be dramatic.

\section{Zero-bias transmission spectra}

\subsection{DTB}

\begin{figure}
\centerline{
\includegraphics[width=7cm]{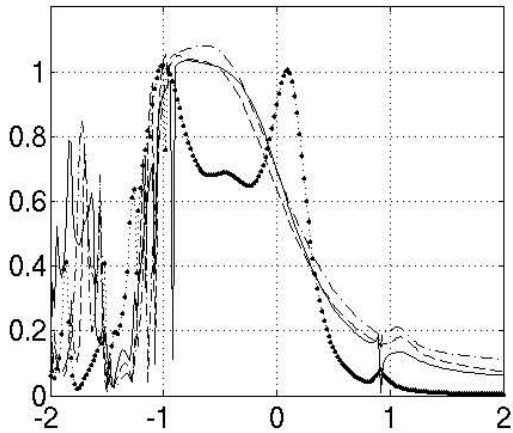}
}
\caption{
Adsorption-site dependence of the transmission $T(E,V_b=0)$ for Au-DTB-Au on flat, relaxed Au(111). Full line: hcp site (3-fold hollow); dashed line: fcc site; dashed-dotted line: bridge site; dotted line: on-top site, not stable, forces on sulfur atom parallel to the gold surface were constrained.
Much of the sharp line structures in this and subsequent figures is due to non-convergence of the matching procedure of the extended molecule in the bulk electrodes; see further Sect. VII.
}
\label{T_DTB_sites}
\end{figure}

Figure \ref{T_DTB_sites} describes the variation of transmission spectra with four different adsorption sites on a flat, relaxed Au(111) surface. The variation of transmission spectra for the stable adsorption sites is very small, which implies that the I-V characteristics of the molecules adsorbed on the flat Au(111) surface at low bias voltage will in practice be indistinguishable. The transmission for the relaxed molecule adsorbed at the hcp site with the benzene ring rotated $90\deg$ is almost indistinguishable from the presented curve (see Sec. VI).

Indeed, if the DTB molecule is relaxed at the cleaved (unrelaxed) Au(111) surface, the result calculated according to our 
setup is also very similar. The results presented by Stokbro {\em et al.} \cite{Stokbro2003} can be distinguished from our data by a more 
pointed and slightly shifted peak between -1 and 0 eV. The reason is 
that the Au-S distance in their setup is smaller by 0.15A and the Au-Au and S-S distances are smaller corespondingly by 0.37A and 0.07A. 
The probable reason is that the relaxation \cite{Stokbro2003} started 
inside the sandwich with some given Au-Au separation, thus modeling 
different experimental situation, as discussed earlier.

For the unstable configurations, the on-bridge adsorbed DTB shows very  similar transmission. Note that for the thinner slab, this site becomes preferable \cite{Nara2004},  and may  also be encountered in dense films as one of the distinct sites in the surface cell  \cite{Krstic2003}. 
However, the transmission spectrum for the on-top configuration is 
quite different, and the broad peak splits into several narrow overlapping 
ones. The corresponding Au-Au separation is at least 0.22A bigger 
than in the other cases and the Au atom under the sulfur is 
additionally pushed from the molecule inside the gold surface by 0.20A, so that the overlap between the gold surface states and the molecular 
states is reduced, reducing the broadening of the transmission 
peaks. 
As a result, one of the transport levels aligns with the Fermi level (Fig.\ref{T_DTB_sites}) the zero bias transmission is enhanced by a factor of two (a similar result is found by Basch et al. \cite{Basch2005}. 

\begin{figure}
\centerline{
\includegraphics[width=8cm]{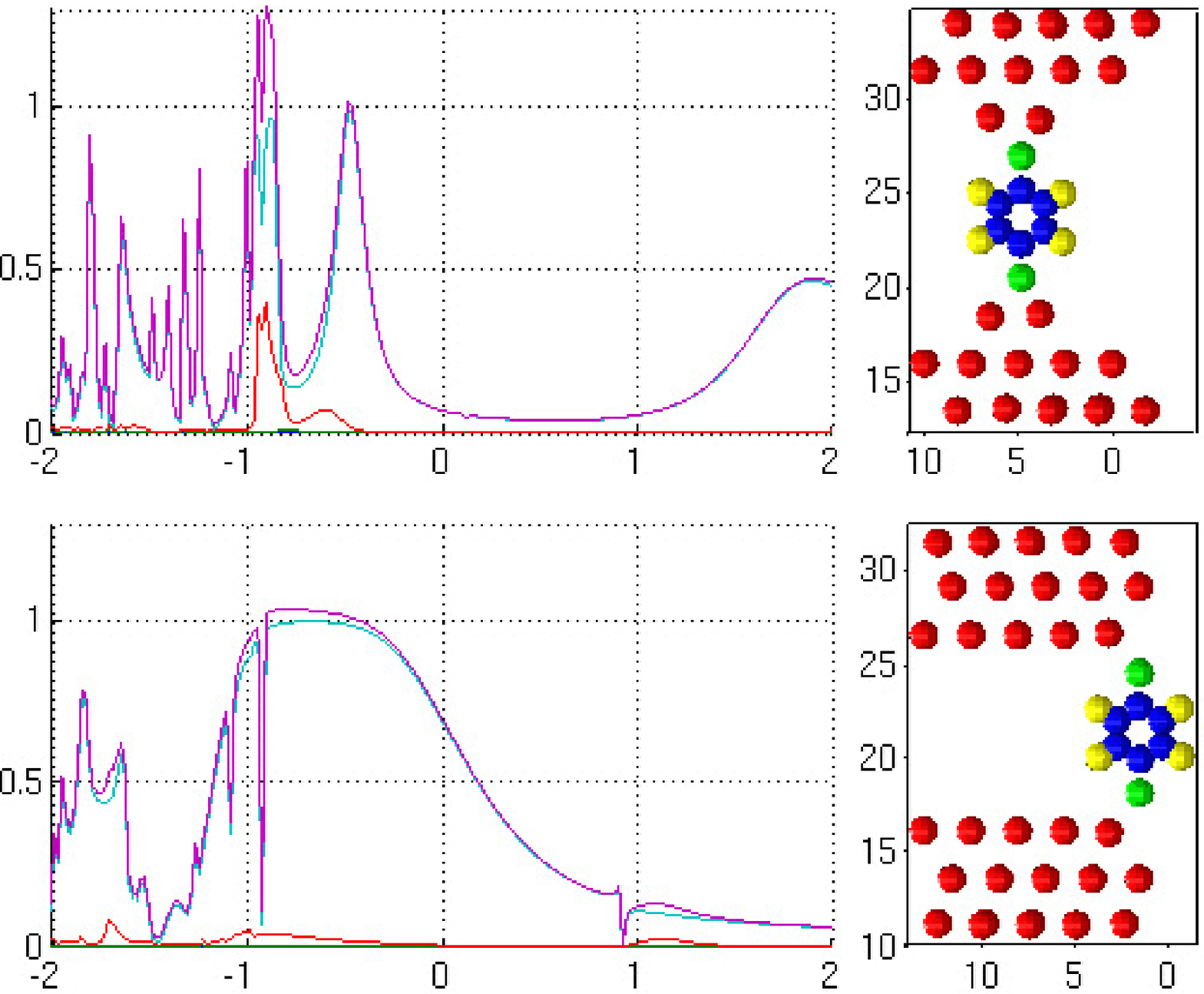}
}
\centerline{
\includegraphics[width=8cm]{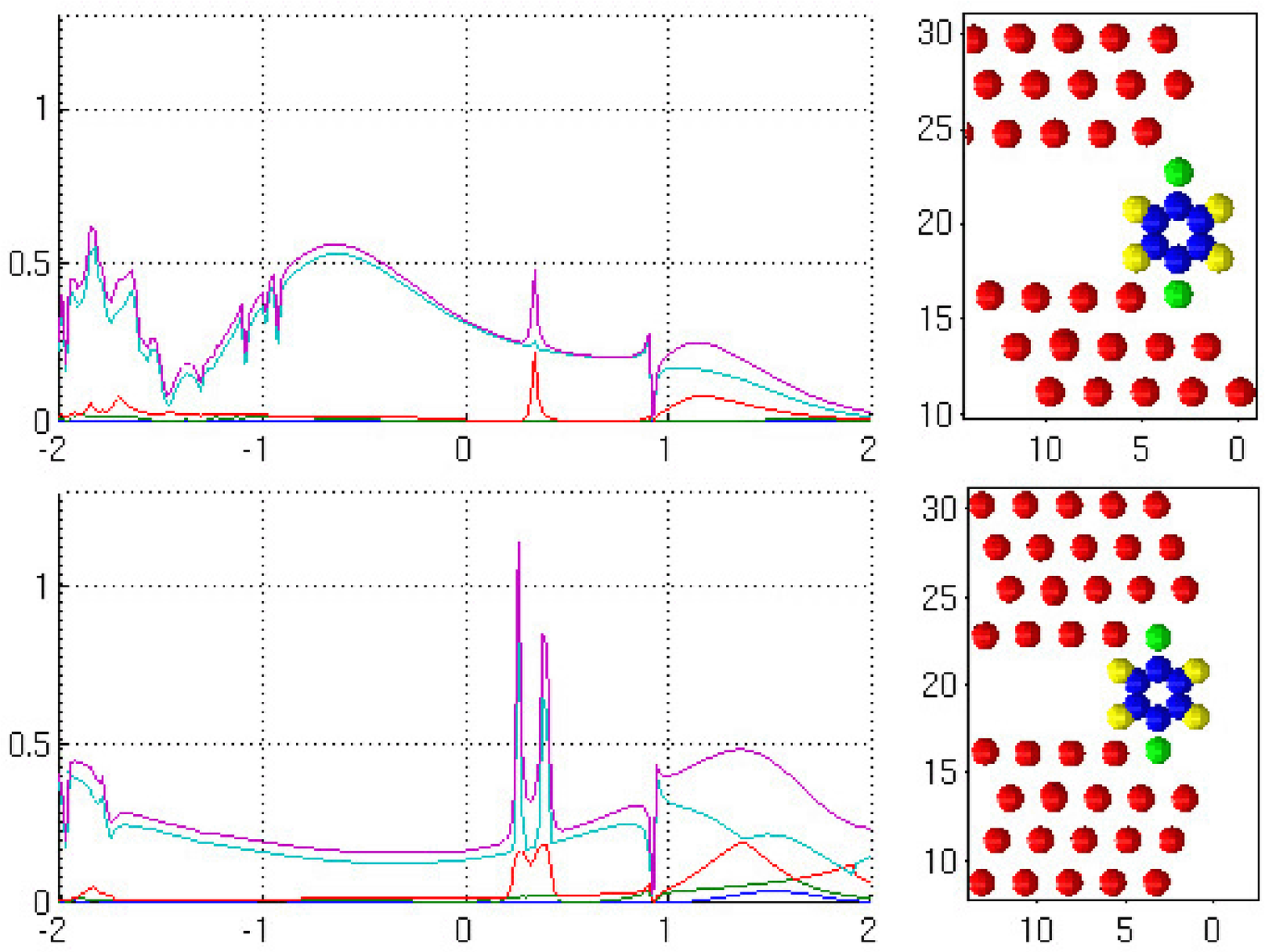}
}
\centerline{
\includegraphics[width=8cm]{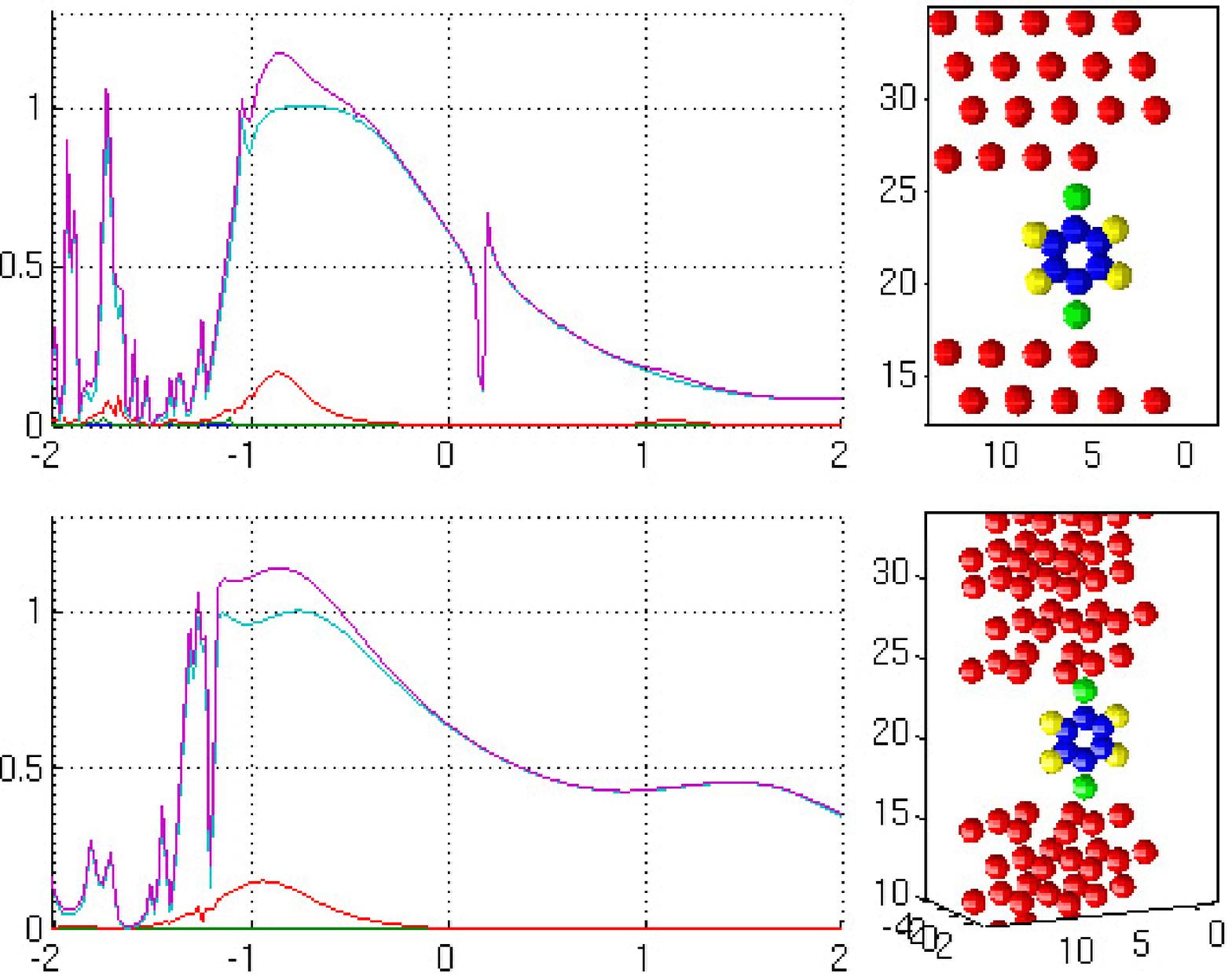}
}
\caption{Adsorption-site dependence of the total transmission $T(E,V_b=0)$ and transmission eigenchannels for DTB in the presence of defects. Top panel: hcp 3-fold hollow on an island of 3 adatoms ("point contact"); second panel: hcp site (3-fold hollow) (flat surface for reference); third panel: hcp site near a vacancy; fourth panel: fcc-bridge site near a vacancy, (meta)stable; fifth panel: one S-atom at flat hcp site, one S-atom buried in vacancy; sixth panel: both S-atoms buried in Au vacancies.} 
\label{T_DTB_contacts}
\end{figure}

Figure \ref{T_DTB_contacts} describes the variation of transmission spectra with lattice defects in Au(111) surfaces, where the S-head is adsorbed on a small island, or near a vacancy, or even in a vacancy.
Proceeding from top to bottom in Fig.\ref{T_DTB_contacts}, a general trend is that the distance between the Au-layers is decreasing. This in itself is a reason for major spectral features of the transmission curves to broaden somewhat, but the main effect is no doubt associated with the variations of the contact structures. We have included our standard DTB spectrum for reference, appearing in the second panel because of the Au-Au layer distance.
In the first panel, the contact has Au(bulk)-Au(island)-S "point-contact" structure, and one may expect that the adsorbate electron density is affected. The quite narrow main HOMO peaks indicate that the corresponding Au-S contact orbitals are more localized than in the DTB reference case (second panel).
Note the similarity with the transmission spectrum of DTB in the "point contact" configuration in Fig.\ref{T_DTB_contacts}. We interpret this as reflecting two separate things: reduced Au-S hybridzation in each contact and reduced hybridization between the two sides (larger electrode-plane/contact separation). 

Note the similarity with the on-top position on the flat Au(111) surface in what concerns the width of the HOMO peaks, indicating weaker coupling of the S-adsorbate to the Au-substrate.

In the next two panels, the S-adsorbate site is close to an Au vacancy (see Fig.\ref{Sites}). The main conclusion is that there are no dramatic changes, in particular not around the Fermi level (the Fano-type resonance in the third panel is spurious; this will be discussed later in Sec. VI). In the fourth panel, the origin of the broad bump above $1 eV$ is not well understood. However, Stokbro {\em et al.} \cite{Stokbro2003} identified a similar, but much weaker, peak in DTB as coming from a single transport channel with orbital distribution similar to the main HOMO-LUMO peaks. A single-channel peak can be sensitive to the particular contact/site configuration and hybridization, which might explain why this peak is prominent in the fourth panel and very weak in panels two and three.

The last two panels involve S-DTB-S adsorbing {\em into} an Au(111) vacancy with one resp. two S-ends. In this case there is a dramatic change of the transmission spectrum and the low-bias conductivity.
We have investigated the bonding structure via the electronic localization function (ELF) and confirmed that the in-vacancy sulfur atom remains bonded to the 
molecule. Mulliken charge analysis also does not show any significant 
change of the charge distribution at the interface. 
However, both the charge density and the ELF  reveal that the bond is 
established between the sulfur and the surface Au atoms, and not to 
the second layer of gold. This significally changes the geometry of the 
interface and might be responsible for significant localization and backscattering in the transport process.

Clearly, with S buried in the Au-surface, the Au:S contact structure is very different. We conclude that the contact states are highly localized in the transport direction and give rise to sharp DOS peaks and very  narrow transport peaks, as seen in the bottom panel in Fig.\ref{T_DTB_contacts}.

We interpret the two sharp peaks above $ E_F $ as reflecting the hybridization between a contact state in each of the two Au-S contacts, resulting in a bonding and an antibonding peak. The fifth panel in Fig.\ref{T_DTB_contacts} then shows the mixture, namely the usual broad HOMO level from the Au-S contact on the flat Au-surface, together with the narrow LUMO peak from the buried Au-S contact. Note that this LUMO peak falls right between the peaks in the next panel, supporting the contact hybridization picture. The fairly large flat spectral density is possibly due to the rather short distance between the Au-electrodes with enhanced non-resonant tunneling through the molecule.

\subsection{DTBd}

\begin{figure}
\centerline{
\includegraphics[width=7cm]{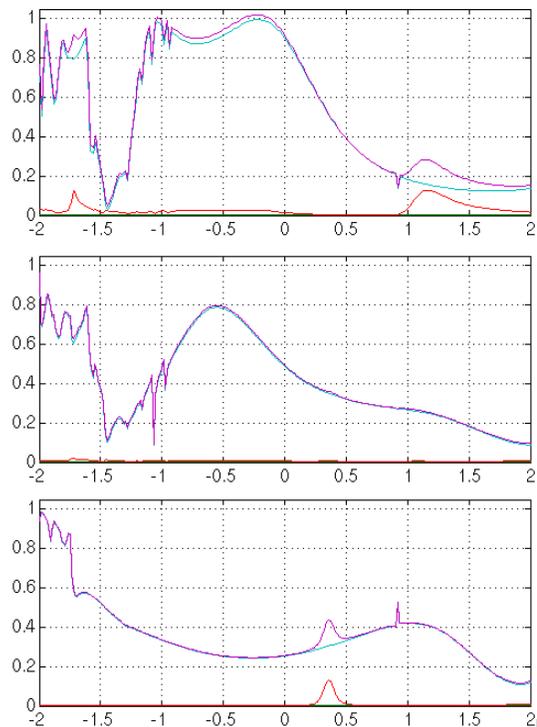}
}
\caption{Total transmission $T(E,V_b=0)$ and transmission eigenchannels for DTBd:  Top panel: sulfur adsorbed at hcp - hcp; middle panel: in-vacancy - hcp; bottom panel: in-vacancy - in-vacancy.}
\label{T_DTBD}
\end{figure}

The zero-bias transmission spectrum of the DTBd molecule (Fig.\ref{4molecules}) - Butadiene-1,4-dithiol - shown in Fig.\ref{T_DTBD} is very similar to that of DTB  in Fig. \ref{4molecules}. This is expected since the $\pi$-bonded benzene ring in DTB corresponds to the straight $\pi$-bonded $C_4H_4$ fragment in DTBd. 
According to the calculation, DTDb is a better conductor and the hybridization betwen the Au-S contacts is stronger than in DTB.

With in-vacancy Au-S contacts (bottom panel) we note, again, that there is a great similarity with DTB. The absence of splitting of the narrow LUMO transport peaks we ascribe to the bigger length (by $0.75$ A) of the DTBd molecule.

\subsection{DTBO}

\begin{figure}
\centerline{
\includegraphics[width=7cm]{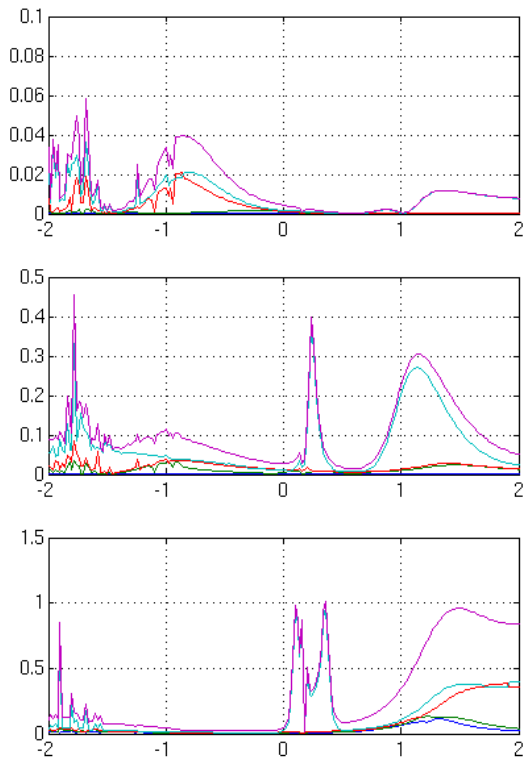}
}
\caption{Zero-bias transmission spectrum $T(E,V_b=0)$ for DTBO; sulfur adsorbed at top: hcp - hcp; middle: vacancy - hcp; bottom: vacancy - vacancy.}
\label{T_DTBO}
\end{figure}

The zero-bias transmission spectrum of 1,3-1,4-dithiol bicyclo[2.2.2]octa-2,5,7-triene (DTBO) (Fig.\ref{4molecules}) is shown in Fig.\ref{T_DTBO}.
DTBO adsorbed on flat relaxed Au(111) is characterized by very poor transmission through the molecule (despite the double-bond structure). The broad Au-S contact structure is visible but only involves a small number of open transport channels with low transmission. However, if the S-contact is buried in an Au vacancy, then the Au-S contact develops a prominent DOS LUMO peak about $0.2eV$ above the Fermi level (Fig.\ref{T_DTBO}, middle panel), as already discussed for DTB and DTBO.

\subsection{DMtB}

The zero-bias transmission spectrum of 1,4 - Benzenedimethanethiol (DMtB), Au-S-CH2-benzene-CH2-S-Au, (Fig.\ref{4molecules}) is shown in Fig. \ref{T_DMtB}:
\begin{figure}
\centerline{
\includegraphics[width=9cm]{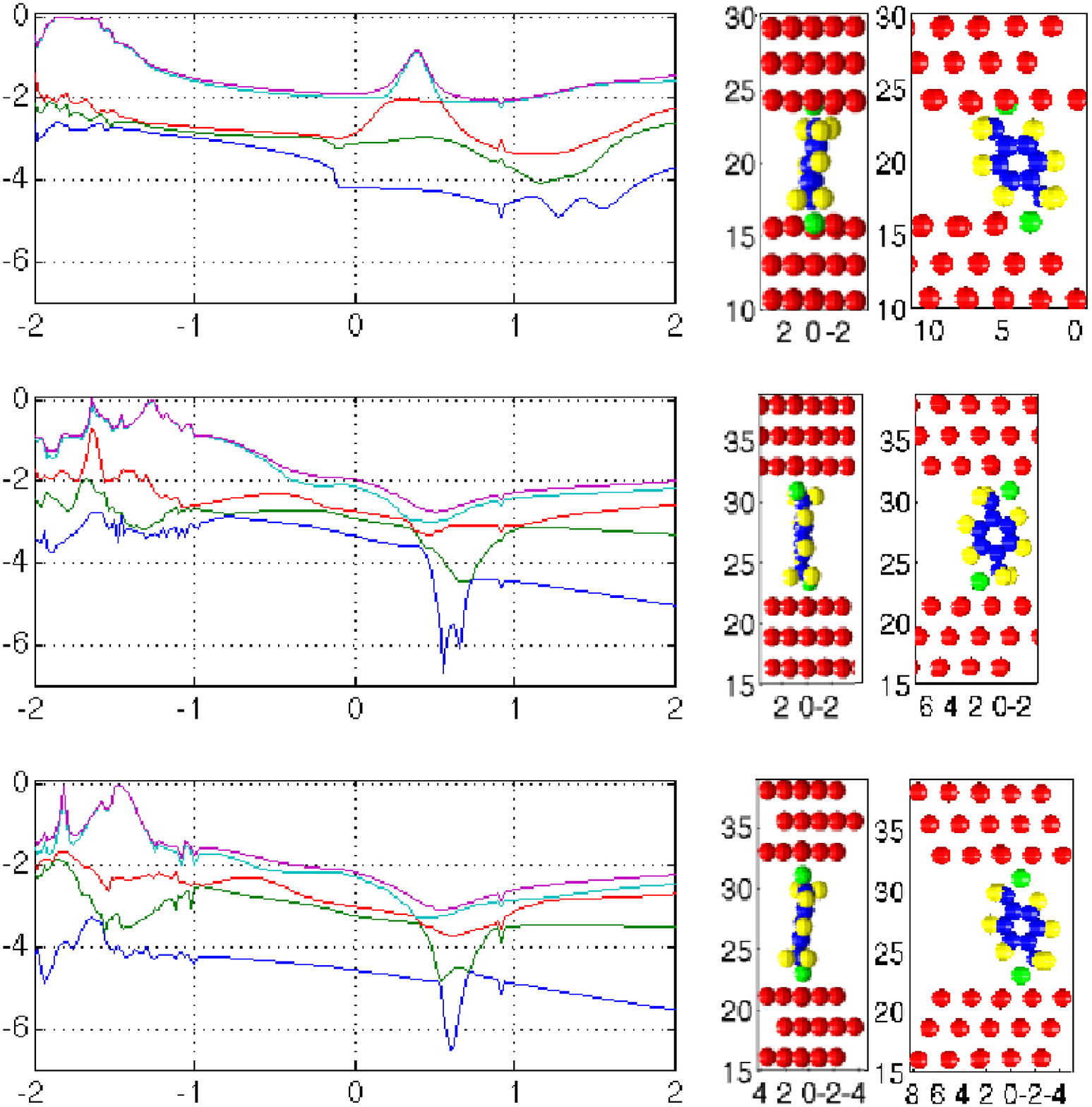}
}
\caption{Left: total transmission $T(E,V_b=0)$ transmission eigenchannels for DMtB, semi-log scale. Right: corresponding setup.}
\label{T_DMtB}
\end{figure}
DMtB presents a case with low transmission through the molecule, and no strong DOS from the contacts. Compared with DTB, the structural  difference is the presence of the sigma-bonded bent CH2 spacers which present big barriers,
leading to very low transmission in a wide region around the Fermi level. 

In Fig.\ref{T_DMtB}, the second and third panels represent Au:S contacts on flat Au(111) surfaces. On the logarithmic scale the contributions from the various channels are all visible, and one finds a prominent (window) LUMO resonance in one of the channels, probably representing resonances in the contacts. 

The first panel shows the result with buried Au:S contacts, and there is now a LUMO resonance structure in one of the channels representing the previosly discussed localized states in these contacts. Again (top panel) there is a flat background and a sharp LUMO peak structure around $0.3 eV$. The difference to the previous cases is that the CH2 spacer barriers now have dramatically lowered the transmission. However, also in this high-resistance case there could be prominent switching behavior.

\section{\label{Transport} Transport properties}

The primary purpose of this work is to investigate the influence of adsorption sites on the transmission properties of M-mol contacts. Even though we draw our conclusions about M-mol-M conductivities mostly from zero-bias transmission spectra, it is extremely valuable to know the M-mol-M transmission as a function of voltage bias in order to understand how levels are pinned to the electrodes and contacts and how the voltage drops are distributed. 

In what follows we will use bias-dependent transmission data to identify specific behavior of  features present in the zero-bias transmission. We have performed a number of self-consistent transmission calculations for bias voltage range close to zero without performing relaxation of the molecular and surface atomic positions. Thus the presented data gives an approximation to the behavior of the transmission spectra in the low-bias regime, which is, however, exact (within the computational method) at zero bias
Therefore we now first discuss how to calculate IVCs, then discuss the general aspects, and finally present results for the bias dependence of transmission spectra for DTB, DTBd and DTBO.

\subsection{I-V characteristics (IVC)}

The nonlinear current through the contact is calculated using the Landauer formula,

\begin{equation}
I(V_{b}) = G_{0}\int^{\mu_{R}}_{\mu_{L}}\, T(E,V_{b})\, dE
\end{equation}
\label{Landauer}

\noindent
where $G_{0}=2e^{2}/h$ is quantum unit of conductance and  $\mu_{R/L} = \pm eV_{b}/2$ are electrochemical potentials of left and right electrode. $eV_b = \mu_R - \mu_L$ is the bias window where left and rightgoing currents balance to give a net transport current $I(V_{b})$ from left to right. The total transmission probability $T(E,V_{b})=\sum_{n=1}^{N}T_{n}(E,V_{b})$ for electrons incident at an energy E through the device under the potential bias $V_{b}$ is composed of all available conduction channels with individual transmission $T_{n}$.
The IVCs can have rather complicated origin because of bias-induced loss of hybridization, or because of both HOMO and LUMO related peaks crossing into the bias window at the same time.

\subsection{General effects of bias voltage}

\begin{figure}
\centerline{
\includegraphics[width=8.5cm]{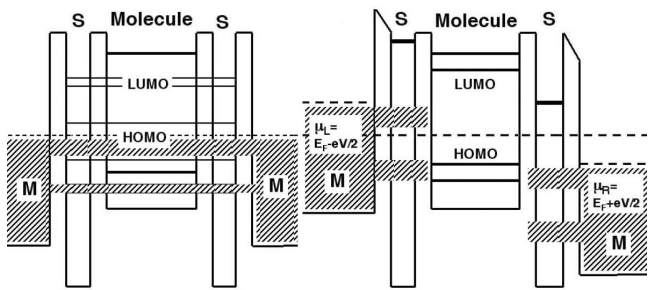}
}
\caption{ 
Schematic quantum well picture illustrating the M-S-mol-S-M system without ($V_b=0$) bias voltage (left) and with ($V_b\ne 0$) bias voltage (right).
}
\label{MBS}
\end{figure}

In simple terms the metal-molecule-metal system can be viewed as a multi-barrier quantum-well structure in between the metal surfaces, as illustrated in Fig.\ref{MBS}.
In this simple picture,  the HOMO-structure of the left electrode will move up toward (and eventually past) the $V_b=0$ "midgap" line, injecting electrons into the right electrode when transport channels are open within the bias window. In the same way, the LUMO structure of the right electrode will move down up toward (and eventually past) the $V=0$ "midgap" line, accepting electrons when transport channels are open within the bias window.

In the case of a symmetric junction (Fig.\ref{MBS}, left), the zero-bias situation involves a maximimum of overlap and hybridization between, in particular, the Au-S contacts, providing high transmission HOMO-LUMO type channels.

When bias is applied (Fig.\ref{MBS}, right), the Fermi levels of the left and right Au-S contacts become displaced by $ \mu_R - \mu_L = eV_b$; the symmetry is broken, lifting the degeneracy and reducing the hybridization. For sufficiently large bias, the left and right contacts become "disconnected" and the level width will reflect that of the isolated Au-S contacts (adsorbate states). In this situation the entire level structure of the left and right Au-S contacts will tend to follow the position of its own Fermi level as it moves with bias voltage. In addition, the molecular structure between the contacts will adjust itself to keep electroneutrality, connecting to one or the other of the contacts, or floating in between. 

As we will see, all of this can introduce various characteristic trends in the bias dependence of the level structures. Level structure that is pinned to the right contact will move down toward negative energies, while structures pinned to the left electrode will move toward positive energies. Levels pinned to both electrodes, or floating, will stay at constant energy independent of bias voltage.

We will recognize these trends in the transmission spectra of DTB, DTBd and DTBO.

\subsection{DTB}

\begin{figure}
\centerline{
\includegraphics[width=8.5cm]{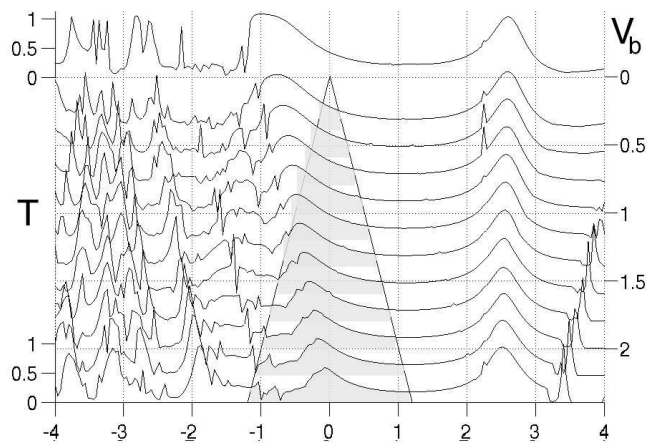}
}
\caption{Bias voltage dependence of the transmission spectrum $T(E,V_b)$ of DTB  adsorbed between hcp sites between flat gold surfaces \cite{Stokbro2005} (previously unpublished). A self-consistent calculation has been employed for each bias voltage. The bias window  $V_b = (\mu_R - \mu_L)/e$ is inside the shaded triangular region. {\em Courtesy of K. Stokbro, 2004}.
}
\label{DTBKurt_biaswin}
\end{figure}

Figure \ref{DTBKurt_biaswin} shows the bias voltage dependence of the transmission spectrum of DTB \cite{Stokbro2003,Stokbro2005}.
The motion of peaks with bias voltage shows a number of clear and simple trends, illustrating clean situations. First of all, most of the HOMO structure appears to move rigidly upwards in energy, pinned to the Fermi level of the left electrode. However, a careful look reveals some deep HOMO structure that moves down pinned to the right electrode. The prominent peak spanning the Fermi level and the bias window is basically pinned to left electrode and strongly narrows and loses intensity under voltage bias due to loss of hybridization upon displacement of the electrodes. On the LUMO side, $E>E_F$, there is one peak clearly moving down pinned (localized) to the right electrode. However, the prominent LUMO level essentially stays put, only moving down very slightly. This LUMO level is delocalized over the entire junction \cite{Stokbro2003} and feels the potential of both electrodes.

Examination of the induced charge distribution and the bias voltage drop across the junction \cite{Stokbro2003} shows that the
potential of the left electrode is fairly constant over the Au-S-B part of the junction and, likewise, for the right electrode and the S-Au contact. The voltage drop occurs at the right contact, between sulfur and the benzene ring. This means, firstly, that the orbitals localized to the  left (Au-S) and right (S-Au) contacts feel a constant potential and can be pinned. Secondly, the injected screening charge should reside in the LUMO, on the benzene ring as close as possible to the right contact. Here it will float in a roughly constant potential. The detailed behaviour in Fig. \ref{DTBKurt_biaswin} (slight LUMO shift to the left) suggests that the potential-drop region moves slightly to the left with increasing bias (so that the LUMO tends to follow the right downgoing ($+V_b/2$) electrode).

The IVC (Stokbro et al.\cite{Stokbro2003}, resulting from integration 
(Eq.(5)) 
over the spectra in Fig.\ref{DTBKurt_biaswin},  displays the generic behaviour of transport through M-mol-M junctions, with linear increase of current with bias voltage, with frequent plateaus, up to a certain bias and then a steep increase at a point where prominent LUMO transmission resonances enter into the bias window.
The low-bias ohmic region of transport follows from any flat density of states around the $V_b=0$ Fermi level, even if the origin is weak tunneling. In contrast to metallic ohmic conductance, which takes place through a conduction band (plane-wave-like states), the M-mol-M junction (flat) density-of-stats is created by overlap of exponentially decaying orbitals and the conductivity (i.e. the slope of the linear region) is exponentially sensitive to the length of the molecule \cite{Crljen2005,Basch2005}.

\begin{figure}
\centerline{
\includegraphics[width=8.5cm]{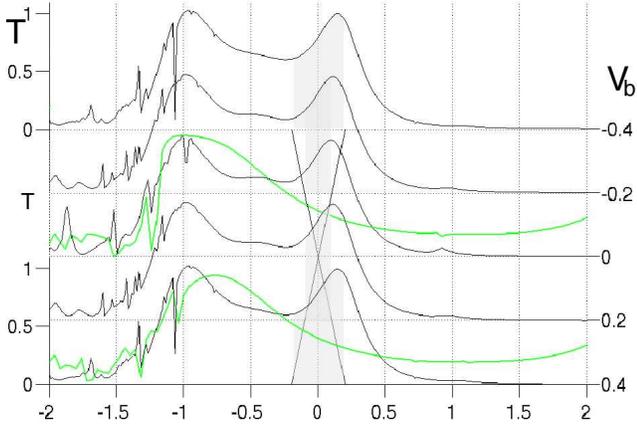}
}
\caption{Bias voltage dependence of the transmission spectrum of $T(E,V_b)$ of DTB adsorbed at the on-top site between flat Au(111) surfaces. The transmission of DTB at the hcp site is shown for reference.}
\label{TEV_DTB_ontop}
\end{figure}

Figure \ref{TEV_DTB_ontop} shows the low-$V_b$ development of the transmission spectrum $T(E,V_b)$ of DTB adsorbed in the on-top position on flat Au(111). In spite of the limited bias range, there is an indication that the prominent peaks move to higher energy,  in a way similar to the broad DTB peak in Fig. \ref{DTBKurt_biaswin}.

\subsection{DTBd and (C$_2$H$_2$)$_n$}

\begin{figure}
\centerline{
\includegraphics[width=8.5cm]{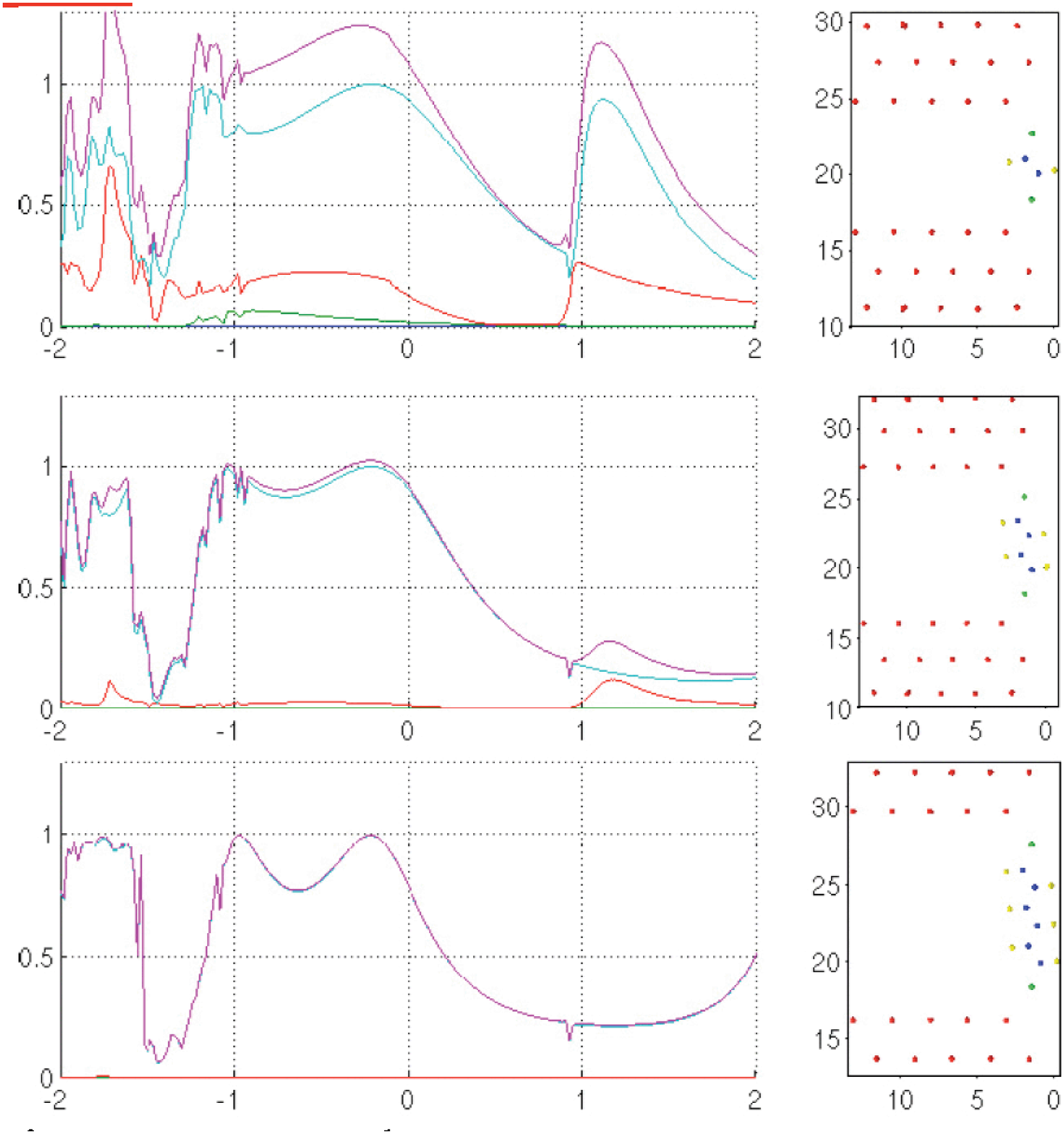}
}
\caption{Zero-bias transmission spectra of Au-S-(CH)$_n$-S-Au for n=1,2,3. DTBd corresponds to n=2.
Note that the issue of chemical stability (for n=1,3) is not addressed (and anyway dependent on the environment).
}
\label{Au-SC6H6S-Au_Vb0}
\end{figure}

Figure \ref{Au-SC6H6S-Au_Vb0} shows the $T(E,V_b=0)$ spectra of Au-S-(CH)$_n$-S-Au for different chain lengths, n=1,2,3, i.e. C$_2$H$_2$, C$_4$H$_4$ (DTBd), and C$_6$H$_6$.
\begin{figure}
\centerline{
\includegraphics[width=8.5cm]{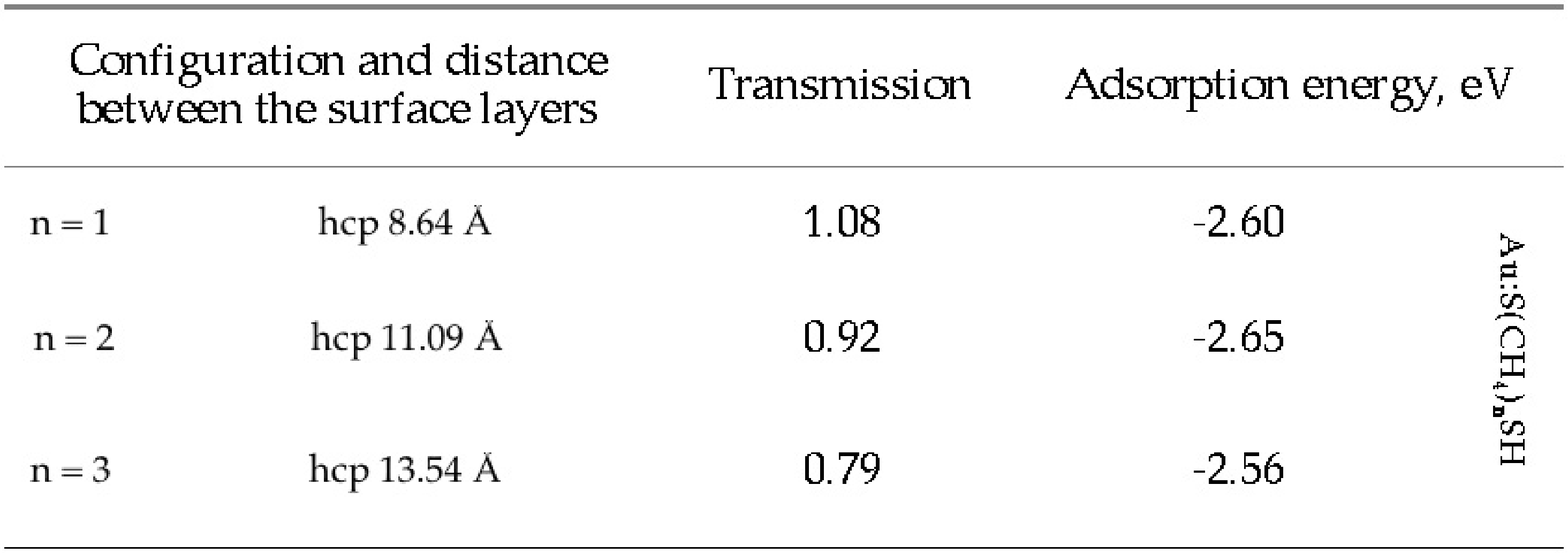}
}
\caption{Table 3: Lengths and zero-bias transmissions of the Au-S-(CH)$_n$-S-Au n=1,2,3 oligomers.}
\label{table3}
\end{figure}
Again, the HOMO peaks narrow up and the LUMO comes down in energy, reflecting reduced Au-S hybridization in each contact and reduced hybridization between the two sides (larger electrode-plane/contact separation). However, the effect is modest because the increase in chain length is small. There is obviously no exponential decrease of the low-bias conductivity in the n=2,3,4 members. However, the narrowing of the peak around the Fermi level in Fig.\ref{Au-SC6H6S-Au_Vb0} suggests that rapid (exponential) decrease of the conductivity may appear already from n=4. Similar behavior was found in the corresponding case of OPVn, for n=3,4,5 by Crljen et al.\cite{Crljen2005}.

An interesting feature is the huge LUMO peak in the 1-1.5 $eV$ range for n=2, essentially wiping out the HOMO-LUMO gap. The peak is also visible for n=4. This peak appears for short electrode distances and is seen also in other systems (e.g., it was discussed in DTB, and was rather prominent for the on-top adsorption site; we mentioned that it seemed to be associated with a rather specific channel, coupling through a LUMO state with orbital character similar to the main HOMO levels just below in energy).

\begin{figure}
\centerline{
\includegraphics[width=8.5cm]{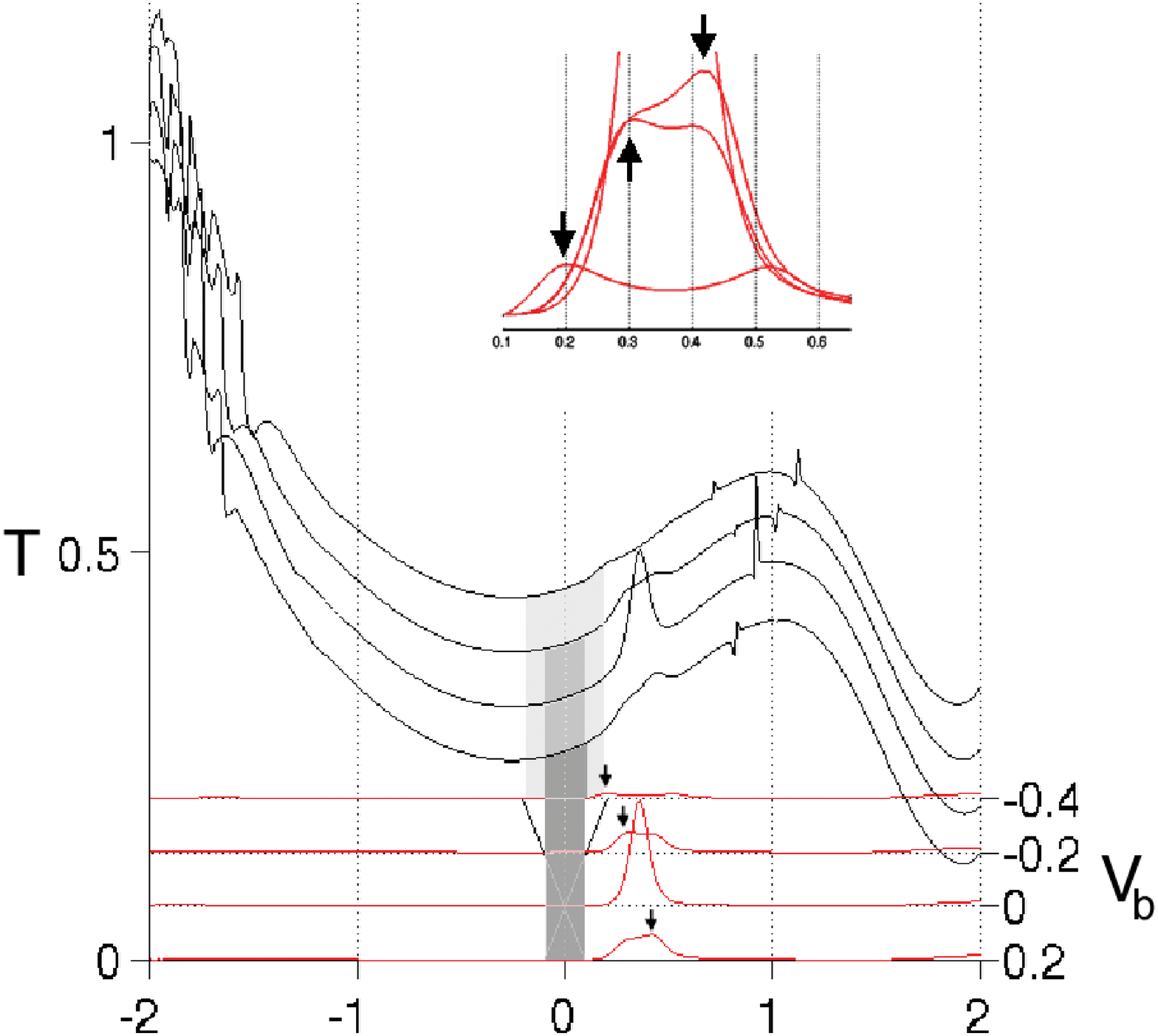}
}
\caption{Bias voltage dependence of the transmission spectrum T(E,Vb) of DTBd adsorbed in vacancy sites. The second transmission channel is shown in red. Arrow marks the higher transmission peak following the left electrode potential.}
\label{DTBd_vacancy}
\end{figure}

Although quite similar, the transmission spectra of DTB (Fig.\ref{T_DTBD}) and DTBd (Fig.\ref{T_DTBO}) with both sulfur atoms adsorbed in Au-vacancies show an interesting difference in what concerns the sharp structure corresponding to the in-vacancy adsorption: there is only one peak visible in DTBd while there are two in DTBO. 
The bias dependent transmission however reveals the two peaks in DTBd. Indeed, as the S-C bond at the left electrode is parallel to the one at the right electrode, while the stacking of the gold layers is reversed, the contacts are not ultimately symmetrical and the peaks can be distinguished, showing that indeed one peak follows the left and another the right electrode potential, see Fig.\ref{DTBd_vacancy}.\\

\subsection{DTBO}

\begin{figure}
\centerline{
\includegraphics[width=8.5cm]{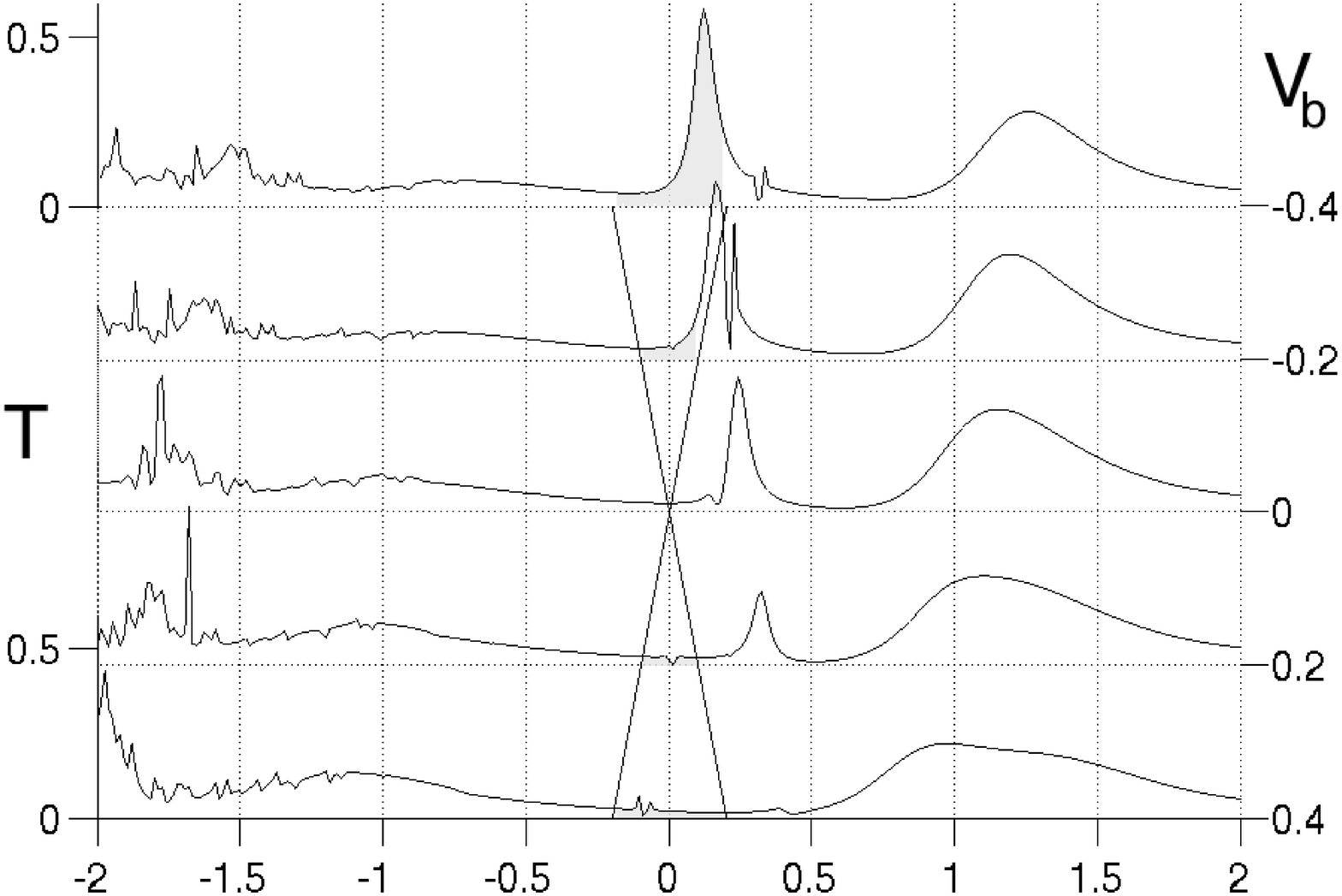}
}
\caption{Bias voltage dependence of the transmission spectrum T(E,V b) of DTBO adsorbed at vacancy Ð hcp sites. 
}
\label{Fig16TVb}
\end{figure}

\begin{figure}
\centerline{
\includegraphics[width=8.5cm]{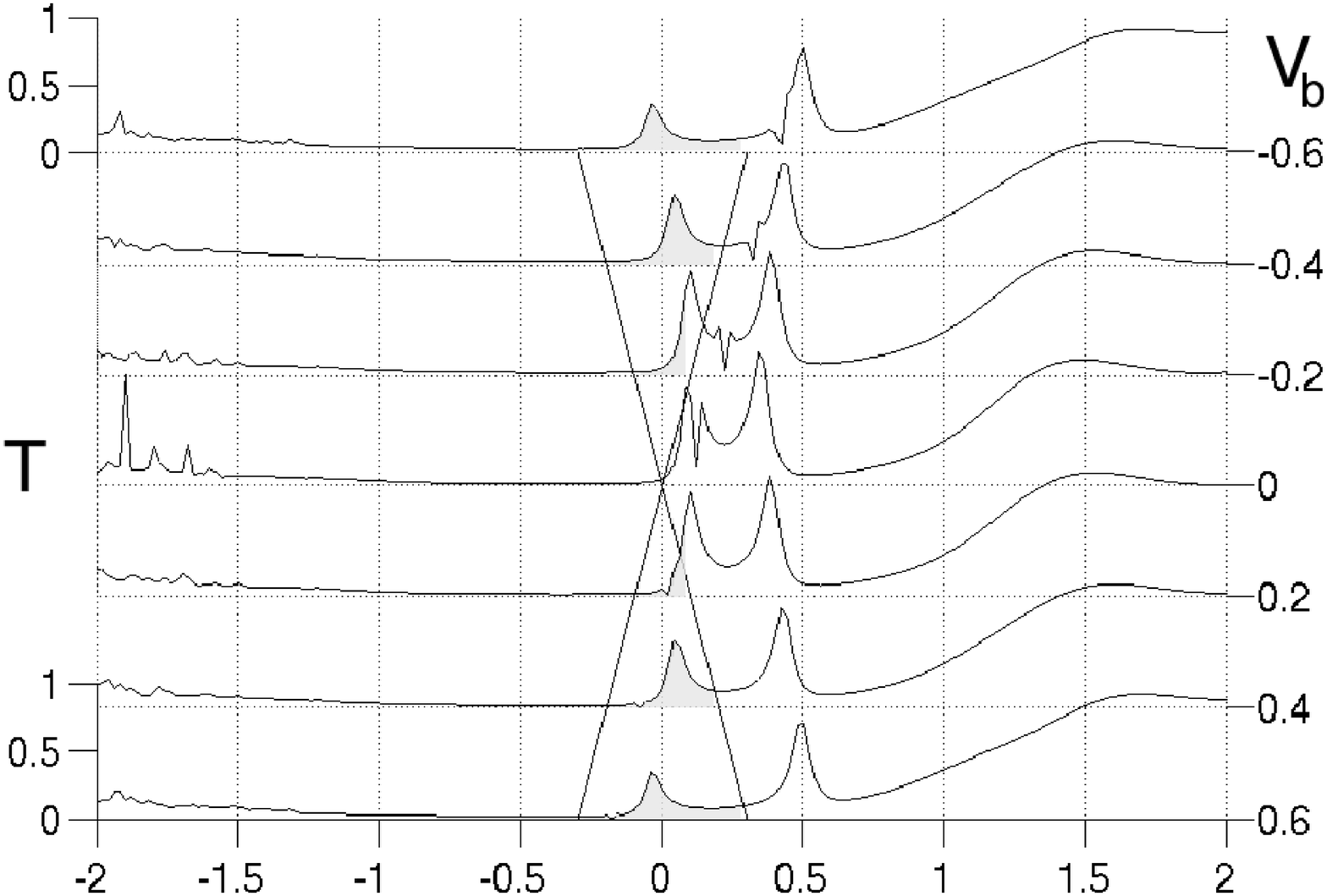}
}
\caption{Bias voltage dependence of the transmission spectrum T(E,V b) of DTBO adsorbed at vacancy Ð vacancy sites.}
\label{Fig15TVb}
\end{figure}

The features introduced by the presence of the surface defects, the most pronounced being pronounced transmission peaks associated with the in-vacancy adsorption of sulfur are especially pronounced in the case of DTBO. We compare the cases with one and two sulfur atoms adsorbed in vacancies in Fig.\ref{Fig16TVb}  and Fig.\ref{Fig15TVb}. 


In general, the transmission peaks follow the corresponding electrode potential.
In Fig. \ref{Fig16TVb}, with only one contact buried (Fig.\ref{T_DTBO}, middle panel), applying positive bias to the Au:S electrode the contact peak  moves toward the zero-bias Fermi level ($E=0$) into the bias window, yielding a high conductivity. (Note, that the broad peak follows the flat (right) electrode potential.)
Reversing the bias, the peak shifts away from $E=0$ {\em and} from the bias window, maintaining the low zero-bias conductivity. 
In the asymmetric case the transmission peak therefore crosses the Fermi level.

On the other hand, in the symmetric case in Fig.\ref{Fig15TVb}, with both contacts buried in vacancies (Fig.\ref{T_DTBO}, bottom panel),  
each contact is associated with a prominent quasi-localized state, and the bonding-antibonding combination (at zero bias) of the symmetric junction leads to a pair of huge LUMO peaks with very big "on-off contrast". Under finite bias voltage,  each peak  follows its own contact. Around zero bias the hybridization of the corresponding transport states results in spectral repulsion, leading to an avoided crossing of the transport peaks (in contrast to DTBd, which remains asymmetric).

\section{Potential molecular switches}

For molecular electronics to be useful one needs to discover or design voltage-controlled switches of electric current. In principle this requires the existence of sharp peaks in the transmission spectrum  $T(E,V_b)$ near the Fermi level that can move into and/or out of the bias window under the influence of source-drain or gate bias voltages.

In order to have a good two-terminal switch, the spectral peaks must be narrow and pinned at the contacts, so that they are easily displaced by small changes of the bias voltage $V_b$. In the case of DTB on-top (Fig.\ref{T_DTB_sites}, top panel) there is a peak at the Fermi level, but the strong hybridization makes it necessary to apply large  $V_b$ to achieve any strong conductance changes. The Au:S "point contact" (Fig.\ref{T_DTB_contacts}, top panel) may be a better candidate. Assuming the prominent peaks to behave like in the $T(E,V_b)$ of DTB (Fig.\ref{DTBKurt_biaswin}), the HOMO should enter the bias window at $|V_b| \sim1 eV$, causing a large and abrupt rise of the conductivity.

As extensively discussed, prominent narrow peaks are however present in the transmission spectra of DTB (Fig.\ref{T_DTB_contacts}) and DTBO (Fig.\ref{T_DTBO}) associated with one or both sulfur heads buried in Au vacancies. The effect is particularly pronounced in DTBO where the background (non-resonant) transmission is very low. 

From the detailed bias-dependence of the DTBO peak  in $T(E,V_b)$ in Fig.\ref{Fig16TVb} we conclude that  this M-DTBO-M junction should behave like a two-terminal rectifying switch, conducting basically only in one bias direction for applied bias voltages exceeding a certain threshold. 

In the same way, from the detailed bias-dependence of the DTBO peaks  in $T(E,V_b)$ Fig. \ref{Fig15TVb},  with both contacts buried in vacancies,  we conclude that the junction should behave like a symmetric bi-directional two-terminal voltage-controlled switch. Below a certain threshold it will be in the off-state, while above this this threshold voltage the junction will be in the on-state in both directions.

Regarding three-terminal transistor switches it should be noted that a third terminal, controlling the (average) potential over the molecular region, will not have any significant effect on the contact states. For a third gate electrode to provide useful transistor action, the molecule itself must provide narrow transmission peaks around the Fermi level, and these states must be only weakly coupled to the contacts. With the systems studied in this work, the best candidate for some significant gate effect should be DTB on-top of an Au island (Fig.\ref{T_DTB_sites}, top panel), where the "point contact" situation to some extent isolates the molecule from the electrodes.

\section{Fano resonances in transmission spectra}

The transmission spectra are built up from a number of transmission resonances involving the states of the junction. The transition amplitude contains a direct path and a path through some localized levels in the junction. If there is a significant direct transmission channel, then the resonance line shapes will show Fano-type asymmetric lineshapes with characteristic interference minima. It should be noted that the "direct" channel could itself be a broad resonance, providing a "non-resonant" background for the resonant level under investigation. Depending on the specific conditions, the Fano resonances can take on a variety of line shapes, from purely symmetric (Breit-Wigner (BW) type, when a direct path is absent, via strongly asymmetric profiles with the interference minimum on either side of the resonance peak, to symmetric window resonances \cite{Fano1961,FanoCooper1965,Wendin1970,Wendin1987}. Figure \ref{DTBKurt_biaswin} provides a large number of examples of such resonances, illustrating all types of line profiles. In particular, one can observe cases where the resonance changes line shape, from BW via asymmetric to window and back to asymmetric, as the resonance moves with bias and the tunneling (hybridization) amplitudes change. 

\begin{figure}
\centerline{
\includegraphics[width=8.5cm]{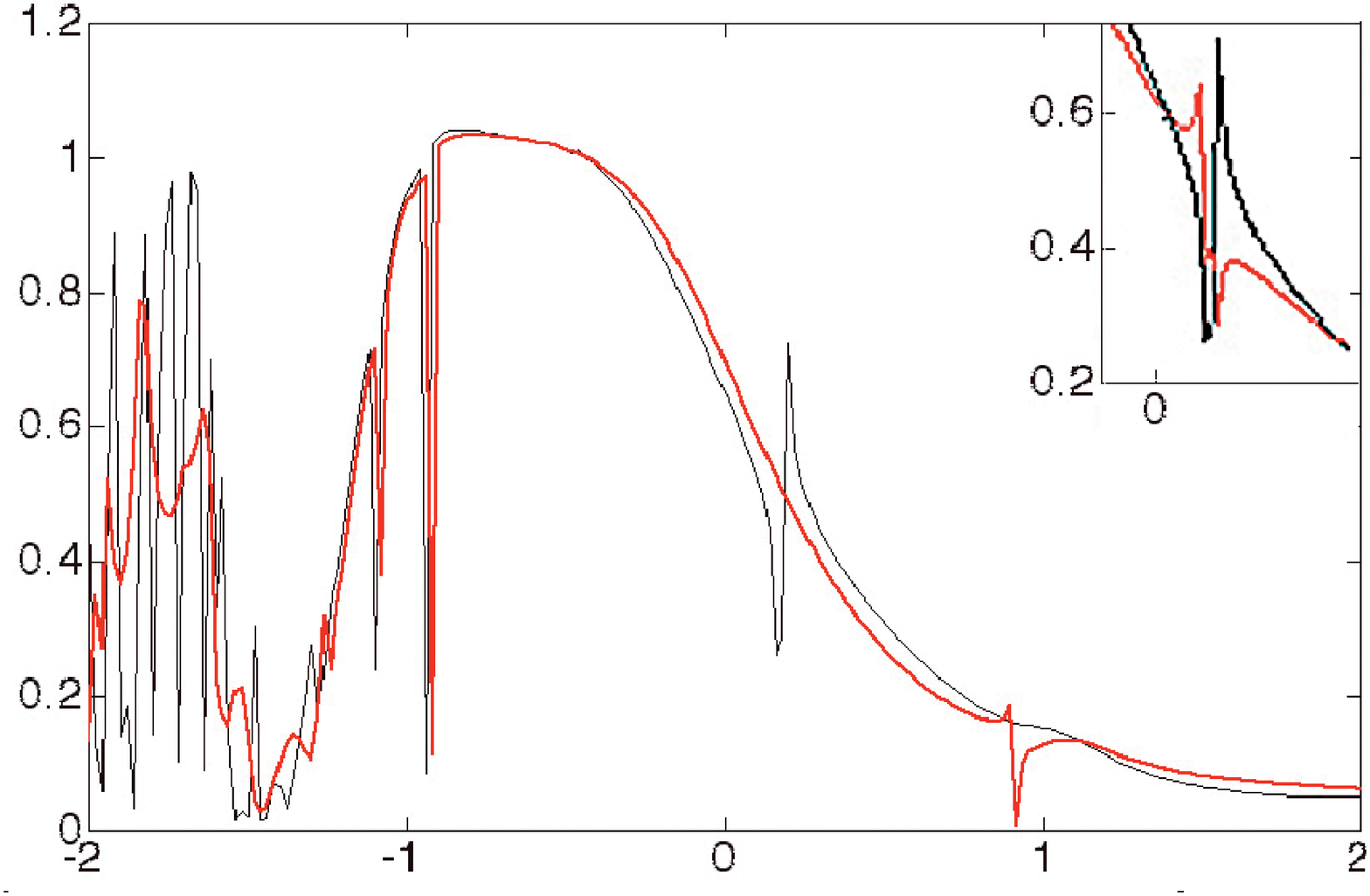}
}
\caption{Transmission T(E, Vb = 0) for DTB with 3x3 (thick line; left Fano resonance) and1x1 (thin line; right Fano resonance) electrodes. Inset: DTB with 3x3 electrodes,the difference between the curves is that the benzene ring has been rotated by 90 deg.}
\label{DTB_Fano}
\end{figure}

Figure \ref{DTB_Fano} gives a specific example of Fano resonances, and at the same time addressing a technical/numerical issue regarding the partitioning of the system into bulk electrodes and a central region. 

The Fano line shapes of transmission resonances tell a lot about the details of hybridization and tunneling elements connecting the  localized state to transport continuum states. As mentioned in Sect. III there is a mathematical boundary separating a bulk electrode from the electrode part that is combined with the molecule, forming the central part of the junction. Since the two regions are described using different types of states (plane waves versus LCAO), the matching procedure represents a critical step. In particular, the cell structure has to be carefully chosen not to give rise to spurious states localized in the vicinity of the mathematical boundary. In Figure \ref{DTB_Fano}, the result of two different choices of cell structure is shown. The resonances around 0.2 eV and 0.9 eV are spurious; however, the inset vividly illustrates the phenomenon of reversal of the line profile due to phase changes in some transport channel.

\section {Discussion}

As mentioned in the introduction, the Au-S-benzene-S-Au junction plays the role of a reference system and has been considered in a number of recent papers. We will therefore compare with a few recent results for zero-bias transmission $T(E)$ spectra,\cite{Bratkovsky2003,Evers2003,Ke2005a,Ke2005b,ThygeJacob2005}  in particular the shapes and widths of $T(E)$ peaks, and the HOMO-LUMO gap and the position of the Fermi level. 
 
A convenient starting point is then the recent work of Ke et al. \cite{Ke2005a} investigating $T(E)$ for Au(111):S in-hollow adsorption for basically two different descriptions of the metal electrodes:  (a) finite-width electrodes and (b) infinite-width supercell periodic arrangement in the x-y directions parallel to the interface. That work demonstrates the appearance of wave-guide resonances in finite-width metal electrodes, giving rise to very pronounced $T(E)$ structure, the conductivity being very sensitive to the exact position of the Fermi level.
However, for infinite (periodic) electrodes in the x-y direction the spectra become dominated by broad peaks very similar to what is shown in our Figs. 8,9,14. 
We conclude that the large width of the HOMO-LUMO peaks basically represents band broadening due to transverse motion (x-y bandstructure) in transmission through the molecular junction, allowing crossing the junction at large angles.
In the present work, this is illustrated in Fig. 9. The second panel is the reference spectrum for Au:S in-hollow adsorption on a flat surface, with a band structure for propagation parallel to the interface, and with a very broad HOMO peak due to a wide cone of transmission angles.

Restricted transverse motion however leads to narrowing of the transmission bands.
In Fig. 9, the top panel describes Au(111):S in-hollow absorption on a 3-atom island, representing a point contact with basically the same Au-S bonding as the flat surface. Our results are similar to the recent results of Thygesen and Jacobsen \cite{ThygeJacob2005}. We can identify two geometric effects: narrowing of the transmission cone and increasing distance between the electrodes. As a result, the HOMO peak structure narrows up, and the strongly Au:S hybridized HOMO-LUMO $"\pi-\pi^*"$ splitting is reduced due to decreased overlap of the Au-S components across the junction. (The LUMO remains rather broad, which can be understood in terms of it being more delocalized). 

Extending to the case of Au:S in the on-top position, the $T(E)$ is shown in Fig. 8, including the flat-surface in-hollow reference spectrum.  In this case the HOMO-LUMO $\pi-\pi^*$ splitting is strongly reduced, which we interpret in terms of strongly decreased Au-S-DTB hybridization. Moreover, the on-top bonding must make the transmission cone more narrow, leading to considerably reduced widths of the HOMO-LUMO peaks. 
Our flat-surface, in-hollow and on-top results in Fig. 8 (and similar results\cite{Evers2003,Ke2005a}) can be directly compared with the results of Bratkovsky and Kornilovitch \cite{Bratkovsky2003}, who find similar $T(E)$  peak positions but considerably more narrow peaks. In line with our discussion above, we believe that this is connected with a more narrow transmission cone (more "one-dimensional" transmission) in their computational scheme,\cite{Bratkovsky2003} possibly due to treating Au-atoms in the Au-S bond separately from the Au-atoms forming the gold electrode. 

A clear difference among several calculations appears to be the position of the Fermi level, which varies by as much as 1 eV between different calculations. 
In the present work, the Fermi level lies in the flank of the HOMO (Figs. 8,9,14), resulting in quite high conductivity, while in Ke et al.\cite{Ke2005a} it lies in the middle of the HOMO-LUMO gap, leading to about a factor of 5 lower conductivity. We note that Ref. \onlinecite{Bratkovsky2003} also places the Fermi level near the middle of the HOMO-LUMO gap, while Refs. \onlinecite{Evers2003,ThygeJacob2005} place it closer to the HOMO.
Differences in the position of the Fermi level indicates different estimates of the charge transfer across the electrode-molecule contacts in equilibrium, connected with general and detailed questions of relaxation and self-consistency. All the quoted papers state that self-consistency and electrode-molecule charge transfer have been taken into account at various levels. 
The accurate placement of the Fermi level remains an important issue in further investigations.
 
At the present levels of accuracy we see no reason for blaming discrepancies between theory and experiment on equilibrium exchange-correlation approximations (EXCA) to DFT (e.g. LDA and GGA). 
The broad HOMO-LUMO peaks, and high conductivity insensitive to adsorption sites and tilting, are connected with idealized situations with short molecules bridging semi-infinite parallel electrodes, allowing transmission within a very wide cone of angles. Such "ideal" junctions are difficult to produce experimentally, and more common situations probably involve connections to non-flat electrodes, via various types of point contacts, chains and wave guides.\cite{Ke2005a,Ke2005b,Basch2005} In these cases the transverse motion may get quantized, creating pronounced transmission structure, and the transmission cone may get narrow, leading to narrow $T(E)$ peaks. One should also mention the possibility of transverse hybridization in densely populated metal-molecule contacts.\cite{Liu2005,PereNewton2005}

For single metal-molecule contacts on semi-infinite electrodes, the 1-dimensional character of the transport will also increase with the length of the molecules due to suppression of large-angle transverse motion across the junction. This is illustrated by the case of Au-S-OPVn-S-Au (n=3,4,5) (having Au-S-benzene contacts) considered by Crljen et al. \cite{Crljen2005}, showing a spectrum of narrow and moderately broad  $T(E)$ peaks. Such systems might be suitable for serious tests of the applicability of DFT (EXCA), the need for time-dependent DFT, or the need for hybrid calculations with better descriptions of diffuse molecular orbitals and correlation effects.

\section {Summary and Conclusions}

Using {\em ab initio} density-functional-based non-equilibrium Green's function methods,
we have studied the bonding-site dependence of the transmission through metal-molecule contacts at Au-S interfaces in molecular junctions of type Au-S-mol-S-Au for a  number of different molecular systems, mainly short double-bond molecules with DTB as a reference system. In some cases the molecules, or the adsorption sites, may be unstable, but the results should nevertheless represent characteristic trends.

We find for all junctions with contacts built on semi-infinite flat Au(111) electrodes that the transmission is rather insensitive to the bonding site, due to the fact that the transmission peaks around the Fermi levels are quite broad.
The exception is (unstable) adsorption on-top of an Au atom, where considerable narrowing of the spectral features near the Fermi level  leads to 30\% increase of the conductivity. More dramatically, if S is adsorbed in an Au vacancy, or on-top of a small (3-Au-atom) island, the transmission can drop very substantially due to mismatch and changes of the level structure in the contacts, allowing order-of-magnitude variations of the conductivity compared with the flat Au(111) surface.

We conclude that the broad transmission peaks can be related to large-angle transverse motion across the junction, and that narrow electrodes, point contacts, impurities, etc., will suppress large-angle transmission. This will break up  the transmission spectrum into a number of narrow peaks, making the transport current much more sensitive to contact sites, tilting and gating.
The sometimes dramatic differences between measured and calculated conductivities may therefore have rather simple explanations.

The present calculations of the bias-voltage dependence of the Au-DTB-Au transmission spectra show that most of the transmission peaks follow the potential of the left or the right electrode.
This can be understood in terms of the Au:S contact states always being pinned to their respective contact due to local charge transfer, following the Fermi level the electrode. A notable exception is the LUMO which is floating at nearly constant energy relative to the zero-bias Fermi level (average electrode potential), in spite of the strong Au:S contact character. We explain this in terms of the LUMO being delocalized over the entire junction, requiring higher bias voltages (than 2.5 Volts) to break the hybridization and disconnect the electrodes.
For symmetric electrodes, at zero bias the contact states are degenerate and we get bonding-antibonding splitting and avoided crossing. At finite bias, the Au:S surface states eventually follow their respective electrode potentials. This behaviour is also evident in Ref. \onlinecite{Crljen2005}, for OPVn junctions with n=3,4,5. 

A potentially very interesting property for molecular electronics is that buried Au-S contacts (S adsorbed in Au vacancy) are associated with very sharp LUMO levels just above the Fermi level. We found this in a number of different systems, and most interestingly in systems with otherwise very low zero-bias transmission at the Fermi level. Possibly such systems, can they be fabricated, will show extremely strong non-linear effects and might work as uni- or bi-directional voltage-controlled 2-terminal switches and non-linear mixing elements. 

\begin{acknowledgments}
We are grateful to Kurt Stokbro for discussion and providing unpublished data, and to Stefan Christiernin, Mark Ratner, Kristian Thygesen and Karsten Jacobsen for reading the manuscript and providing many helpful comments. This work was supported by the IST-FET-NANOMOL project of the EC, by the Swedish Research Council, by the Swedish Strategic Research Foundation, and the Ministry of Science and Technology of the Republic of Croatia.
\end{acknowledgments}


\begin{thebibliography}{00}
\bibitem{SIESTA} 
A double $\zeta +$ polarization basis for the opvn molecule,
while a double $\zeta$ for the gold {\it s} channel and single $\zeta$ for the gold
{\it p} and {\it d} channel was used.
\end{thebibliography}
\end{document}